\documentclass[twocolumn,pra,amsmath,amssymb,superscriptaddress,longbibliography]{revtex4-2}

\pdfoutput=1
\usepackage{hyperref}
\usepackage{graphicx}			
\usepackage[T1]{fontenc}
\usepackage{color}
\usepackage{mathtools}
\usepackage[normalem]{ulem}
\newcommand{\ket}[1]{\left| #1 \right>} 
\newcommand{\bra}[1]{\left< #1 \right|}

\renewcommand{\Im}{\operatorname{Im}}
\newcommand{\rmi}{\mathrm{i}}
\newcommand{\rmd}{\mathrm{d}}
\newcommand{\eps}{\varepsilon}
\newcommand{\rme}{\mathrm{e}}

\newcommand{\Log}{\mathrm{Log}}

\usepackage{framed}
\definecolor{shadecolor}{gray}{0.95}
\hypersetup{
    colorlinks=true,
    linkcolor=black,
    citecolor=black,
    filecolor=black,
    urlcolor=blue,
}
\newtheorem{theorem}{Theorem}
\newtheorem{proposition}{Proposition}
\newtheorem{conjecture}{Conjecture}
\newtheorem{lemma}{Lemma}

\begin{document}
\date{\today}							
\title{Bulk-boundary correspondence and singularity-filling in long-range free-fermion chains}

\author{Nick G. Jones}
\affiliation{Mathematical Institute, University of Oxford, Oxford, OX2 6GG, UK}
\affiliation{The Heilbronn Institute for Mathematical Research, Bristol, UK}

\author{Ryan Thorngren}
\affiliation{Kavli Institute of Theoretical Physics, University of California, Santa Barbara, California 93106, USA}

\author{Ruben Verresen}
\affiliation{Department of Physics, Harvard University, Cambridge, MA 02138, USA}

\begin{abstract}
The bulk-boundary correspondence relates topologically-protected edge modes to bulk topological invariants, and is well-understood for short-range free-fermion chains. Although case studies have considered long-range Hamiltonians whose couplings decay with a power-law exponent $\alpha$, there has been no systematic study for a free-fermion symmetry class. We introduce a technique for solving gapped, translationally invariant models in the 1D BDI and AIII symmetry classes with $\alpha>1$, linking together the quantized winding invariant, bulk topological string-order parameters and a complete solution of the edge modes.  
The physics of these chains is elucidated by studying a complex function determined by the couplings of the Hamiltonian: in contrast to the short-range case where edge modes are associated to roots of this function, we find that they are now associated to singularities.
A remarkable consequence is that the finite-size splitting of the edge modes depends on the topological winding number, which can be used as a probe of the latter.
We furthermore generalise these results by (i) identifying a family of BDI chains with $\alpha<1$ where our results still hold, and (ii) showing that gapless symmetry-protected topological chains can have topological invariants and edge modes when $\alpha -1$ exceeds the dynamical critical exponent.
\end{abstract}
\maketitle
\textbf{Introduction.}
The bulk-boundary correspondence is a central concept in the study of topological phases of matter \cite{Motrunich01,Ryu02,Li08,Schnyder08,Schnyder09,Hatsugai09,Hasan10,Ryu10,Delplace11,Mong11,Tanaka12,Graf13,Bernevig15,Asboth16,Peng17,Sedlmayr17,Rhim18}. This relates topologically stable edge effects with topological features of the bulk Hamiltonian. A simple manifestation of this is in certain translation-invariant quantum chains with time-reversal symmetry, where the Hamiltonian on a periodic chain can be used to define a winding number which counts the number of topologically protected Majorana zero modes localised at the edge \cite{Motrunich01,Schnyder08,Kitaev09,Fidkowski10,DeGottardi13,Verresen18}. Research on this topic has predominantly focused on the short-range case where lattice Hamiltonians couple sites up to some finite range.
In the past decade there has been significant interest in quantum systems with long-range interactions \cite{Maity19,Defenu21}. This has been motivated by proposals for, and progress in, experimental systems, such as Ref.~\cite{Pientka13} for effective free-fermion chains. 
Here long-range typically means that couplings decay as a power of the distance [i.e., Hamiltonian terms acting between sites at distance $r$ are $  O(r^{-\alpha}$)].
Interesting physical effects have been observed including algebraically localised edge modes and the breakdown of the entanglement area law \cite{Gong17} and conformal symmetry at criticality \cite{Lepori16}.

Regarding topological edge modes in such long-range chains, most results in the literature concern the canonical Kitaev chain \cite{Kitaev01} with additional long-range hopping or pairing terms \cite{Vodola14,Vodola15,Lepori17,Alecce17,Patrick17,Maity19,Jaeger20,Kartik21,Sadhukhan21,Francica22}. (For interacting studies see Refs.~\cite{Gong16,Lapa21}.) 
The long-range Kitaev chain sits in the \emph{BDI symmetry class} of free-fermion Hamiltonians \cite{Altland97,Schnyder08,Kitaev09,Ryu10}, and it is straightforward to see that for $\alpha>1$ the bulk winding number remains well defined \cite{Lepori17}. Very recently, Ref.~\cite{Gong22} treated the free-fermionic phase diagram in great generality and gave a proof that the short-range phase classification is preserved in the long-range case with $\alpha>d$ (in general dimension and symmetry class).
Work on the long-range Kitaev chain showed that topological edge modes exist, but only in particular models. This leaves open important questions for topological Majorana zero modes in long-range chains: when do they exist, what is their connection to the bulk invariant, and what are their localisation properties at the edge?

Here, we present the first systematic study of a whole symmetry class, giving rise to a detailed bulk-boundary correspondence in long-range chains.
We focus on the exemplary BDI class as mentioned above, although the results carry over for the AIII class \footnote{While this is a physically different setting, the analysis is almost identical, see Appendix \ref{app:AIII}.} which famously includes the Su-Schrieffer-Heeger chain \cite{Su79}.

We show that the bulk invariant corresponds exactly to the number of topological edge modes and give a rigorous method to find the edge mode wavefunctions. Additionally, we find that the bulk string-order parameters for the short-range case continue to reveal the bulk topology. We complement 
these results by outlining a principle for calculating the finite-size energy splittings for the zero modes in long-range chains, that we call \emph{singularity filling}. 
Together with our analysis of the localisation properties of the edge modes, this brings a number of disparate results in the literature into a coherent picture.

The methods we use are from the mathematical theory of Toeplitz determinants (see, e.g., \cite{Boettcher06}), a key technique in the analysis of the two-dimensional Ising model \cite{McCoy73}. We expect this approach to long-range chains to be fruitful more generally. 

We use the standard notation $g(n)=O(h(n))$ when $ g(n) \leq \textrm{const}\times h(n)$ for $n$ sufficiently large, and $g(n)=\Theta(h(n))$ when $g(n)=O(h(n))$ and $h(n)=O(g(n))$.

\textbf{The model.}
Consider the BDI class of translation invariant spinless free-fermions with time-reversal symmetry:
\begin{align}
H_{\mathrm{BDI}} = \frac{\rmi}{2}  \sum_{m,n \in \textrm{sites}} t_{m-n} \tilde \gamma_n \gamma_{m}. \label{eq:BDI} \end{align}
Here $\gamma_n = c_n^{\vphantom{\dagger}}+c^\dagger_n$ [$\tilde\gamma_n = \rmi( c_n^{\dagger}-c^{\vphantom{\dagger}}_n) $] are the real [imaginary] Majorana fermions constructed from spinless complex fermionic modes $c_n$ on each site.
The real coupling coefficients $t_n$ are called $\alpha$-decaying \cite{Gong22} if $t_n  \leq \textrm{const} (1+\lvert n\rvert)^{-\alpha}$. 
Assuming absolute-summability of the $t_n$ (implied by $\alpha>1$) we can solve the closed chain by a Fourier transformation and Bogoliubov rotation
(see Appendix \ref{app:solution}). This information is summarised by the continuous complex function:
\begin{align}
f(z) = \sum_{n=-\infty}^\infty t_n z^n, \qquad z = \rme^{\rmi k} \quad 0\leq k< 2\pi. \label{eq:fz}
\end{align} 
The eigenmode with momentum $k$ is defined by the phase of  $f(\rme^{\rmi k})$ and has energy $\eps_k = \lvert f(\rme^{\rmi k})\rvert$. Thus, the Hamiltonian \eqref{eq:BDI} is gapped when $f(z) \neq 0$ on the unit circle. In that case, the argument of $f(z)$ is well-defined, and we have the winding number \begin{align}
\omega = \lim_{\eps\rightarrow 0}\left( \textrm{arg}(f(\rme^{\rmi (2\pi-\eps)}))- \textrm{arg}(f(\rme^{\rmi \eps})) \right) \in \mathbb{Z}.
\end{align}
This is the \emph{bulk topological invariant}, which cannot change without a gap-closing if we enforce the absolute-summability condition. 

\textbf{Bulk-boundary correspondence and edge mode wavefunction.}
We now consider the Hamiltonian \eqref{eq:BDI} with open boundary conditions (we keep only the couplings that do not cross the boundary). We first consider the limit of a half-infinite chain, where edge modes have zero energy (later we study finite-size splitting).

In this limit, the edge mode wavefunctions are zero-eigenvectors of a \emph{Toeplitz operator}, which can be solved using the \emph{Wiener-Hopf method}. More directly, define a real Majorana zero mode as $\gamma_L = \sum_{n=0}^\infty g_n \gamma_n$ that satisfies $[\gamma_L, H_{\textrm{BDI}}]=0$. 
Evaluating the commutator gives us a Wiener-Hopf sum equation, which is straightforwardly solved \footnote{The proof is given in Appendix \ref{app:Wiener-Hopf}. To exclude accidental edge modes we need results found in Refs.~\cite{Widom60,Douglas80,Boettcher06}.} using results of McCoy and Wu \cite{McCoy73},
leading to:
\begin{theorem}[Bulk-boundary correspondence]\label{thm1}
Take a half-infinite open chain $H_\textrm{BDI}$, where the related bulk Hamiltonian has winding number $\omega$ and absolutely-summable couplings, then there exist exactly $\lvert\omega\rvert$ zero-energy edge modes. 

More constructively, writing $f(z) = z^\omega b_+(z)b_-(z)$ (here $b_\pm(z)$ are the Wiener-Hopf factors defined below), then for $\omega>0$ we have $\omega$ linearly independent normalisable real edge modes given by $\gamma_L^{(m)}= \sum_{n=0}^\infty g_{n}^{(m)} \gamma_n$ with $g_n^{(m)}= (b_-(1/z)^{-1})_{n-m}$ for $0\leq m \leq \omega-1$.
  
For $\omega<0$ the same results hold upon substituting $\gamma_n \to \tilde \gamma_n$ and $b_-(1/z)^{-1} \rightarrow b_+(z)^{-1}$.
\end{theorem}
Here and throughout we use the notation that $(h(z))_n ={(2\pi\rmi) }^{-1} \int_{S^1} h(z) z^{-(n+1)}\rmd z$ is the $n$th Fourier coefficient of a function $h(z)$. 
Key to our result is a canonical form called the Wiener-Hopf decomposition.
First define $f_0(z) = z^{-\omega}f(z)$, which is non-vanishing on the unit circle and has a continuous logarithm $\log(f_0)(z)$.
We fix the normalisation of $H_\textrm{BDI}$ such that the zeroth Fourier coefficient $(\log(f_0))_0=0$.
Then we can always write:
\begin{align}
f(z) = z^\omega \; b_+(z)\; b_-(z), \label{eq:WH}
\end{align}
where the Wiener-Hopf factor given by $b_\pm(z) = \rme^{\sum_{n=1}^\infty (\log(f_0))_{\pm n} z^{\pm n}}$ is analytic strictly inside (outside) the unit disk. We note that $z^{\omega}$ encodes the winding around the unit circle and hence the topological invariant of the system. Multiplying $f(z)$ by $z^{m}$ shifts \footnote{This shift is an application of an `SPT entangler' \cite{Jones21a,Tantivasadakarn23}.} the hopping $t_n \to t_{n-m}$, such that $f_0(z)$ defines a topologically trivial `version' of the system. This is analogous to the trivial insulator and the Kitaev chain being related by a shift.

Theorem \ref{thm1} extends the bulk-boundary correspondence from the short-range to the long-range case: the bulk winding number counts edge modes everywhere in the space of Hamiltonians with absolutely-summable couplings [$(\alpha>1)$-decay implies absolute-summability, but examples like the Weierstrass function \cite{Zygmund02,Grafakos08} can be used to construct families with $0<\alpha\leq 1$]. Our result is also constructive: we have the edge mode wavefunction in terms of Fourier coefficients of a particular function. To construct the exact edge mode, one needs to first calculate the Wiener-Hopf decomposition. However, we will see below that this can often be bypassed if one is interested only in the asymptotic edge-mode profile.

In short-range models we expect exponentially-localised edge modes, corresponding to roots of $f(z)$ \cite{Verresen18,Balabanov21} (see Appendix \ref{app:shortrange}). Based on Theorem \ref{thm1} we see that the localisation follows from analytic properties of the Wiener-Hopf factors. If $(b_\pm(z^{\pm1}))^{-1}$ is analytic to some distance outside the unit circle, we will see exponential decay (this appears consistent with previous such observations in the long-range Kitaev chain at fine-tuned points \cite{Jaeger20}). Exponential localisation was also observed in Ref.~\cite{Lapa21}, but for a different reason---there the short-range (parity-odd) edge modes cannot couple to the long-range density-density interactions in perturbation theory due to fermion parity symmetry. In our long-range case, the edge modes are generically algebraically-decaying and guaranteed to be normalisable due to the Wiener-L\'evy theorem \cite{McCoy73,Zygmund02}.

\textbf{Example.}
Consider $f(z) = z^\omega \textrm{Li}_\alpha(z) \textrm{Li}_\alpha(1/z)$, where $ \textrm{Li}_\alpha(z)= \sum_{k=1}^\infty z^k/k^\alpha$ is the polylogarithm of order $\alpha>1$. 
The couplings $t_n$ are $\alpha$-decaying and moreover $t_n = \Theta(n^{-\alpha})$ for $n\rightarrow \pm\infty$.

One can read off $b_+(z) = \textrm{Li}_\alpha(z)/z$ and $b_-(z) = z\textrm{Li}_\alpha(1/z)$. Suppose $\omega =1$, then we have one 
edge mode with:
\begin{align}
g_n = \frac{1}{2\pi \rmi}\int_{S^1} \frac{z^{-n}}{\textrm{Li}_\alpha(z)}\rmd z = - \frac{1}{\zeta(\alpha)^2 n^{\alpha}}(1+o(1));\label{eq:example1}
\end{align}
the second equality is derived using contour integration and known asymptotics for $\textrm{Li}_\alpha(z)$ on the real line (assuming $\alpha\notin\mathbb{N}$) \cite{NIST:DLMF}. The analysis is given in Appendix \ref{app:example} and further terms in the asymptotic expansion can be found using the same methods.

For $\omega =2$, we see we have two edge modes, with the same leading order behaviour. This means we can take the difference $n^{-\alpha} - (n-1)^{-\alpha}= \Theta (n^{-\alpha-1})$, and have a faster decaying strictly localised mode (see Theorem \ref{thm2}).

\textbf{Singularity-filling for wavefunctions.}
\begin{figure}
\includegraphics[]{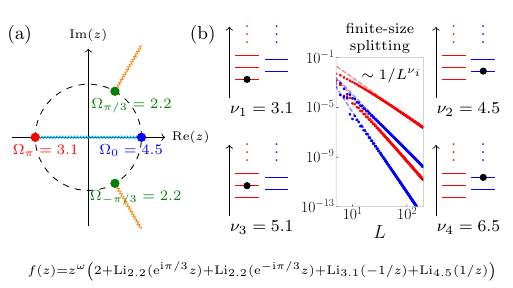}
\caption{\textbf{Finite-size splitting from singularities.} (a) As an example of our general results, we consider a long-range chain whose hopping coefficients define the complex function $f(z)$ [Eq.~\eqref{eq:fz}] with singularities of $f(1/z)^{-1}$ depicted. According to Conjecture \ref{conjecture}, the power-law exponents associated to these singularities dictate the finite-size energy splitting of the $\lvert \omega \rvert$ Majorana edge modes. (b) We illustrate this for $\omega=4$, where we show the numerically-obtained splittings for system size $L$. Their power-law decays $\sim1/L^{\nu_i}$ are accurately predicted by the `singularity-filling' of Conjecture \ref{conjecture}. For $\omega>0$ the singularities associated to branch cuts inside the unit disk matter [i.e., $\Omega_0 = 4.5$ (blue) and $\Omega_\pi = 3.1$ (red)]; for $\omega<0$ this is reversed (see Appendix \ref{app:numerics}). \label{fig:1}}
\end{figure}
While the bulk-boundary correspondence of Theorem \ref{thm1} is our most general result, we can give additional results in a broad class of $(\alpha>1)$-decaying models.
We say that $1/f(1/z)$ has singularities at $\{k_s\}_{1\leq s\leq r}$ if it has asymptotic Fourier coefficients 
$(1/f(1/z))_{n} =\sum_{s=1}^{r} \rme^{\rmi n k_s }n^{-\Omega_{k_s}} \left( a_s + o(1) \right)$ as $n\rightarrow + \infty$.
 We call $\Omega_{k_s}>1$ the order of the singularity at $k_s$, and assume the $o(1)$ term is `nice', i.e., can be expressed as a sum of inverse powers of $n$ [as is the case in Eq.~\eqref{eq:example1}]. We also assume that $\Omega_{\textrm{min}}=\min_s\{{\Omega_{k_s}}\}\notin\mathbb{Z}$. This implies that $1/f(1/z)$ has $\delta_0=\lfloor \Omega_{\textrm{min}} -1\rfloor $ continuous derivatives \cite{Boettcher06b,Grafakos08}.
\begin{theorem}[Edge mode from singularity-filling]\label{thm2}
Consider the set-up as in Theorem \ref{thm1} with $\omega>0$, and suppose in addition that $1/f(1/z)$ has singularities as defined above. Define $\nu_1,\dots,\nu_\omega$ by the $\omega$ lowest levels $\mathcal{E}_s(n) = \Omega_{k_s} + n$ over all singularities $s$ and $n\in \mathbb{Z}_{\geq0}$ (`singularity filling') and define $\nu_{\star} = \delta_0+\Omega_{\textrm{min}} -1$.

We can find a basis of mutually anticommuting edge modes
$\hat{\gamma}_L^{(p)}=\sum_{n=0}^\infty \hat{g}_n^{(p)} \gamma_n$
where $ \hat{g}_n^{(p)}=O(n^{-\tilde\nu_p})$;
for $\tilde\nu_p= \min\{\nu_p,\nu_\star\}$.

For $\omega<0$ analogous results hold where  we now take $\gamma_n\rightarrow \tilde\gamma_n$ and $f(1/z)\rightarrow f(z)$.
\end{theorem}
The idea of the proof is as in the $\omega=2$ example following \eqref{eq:example1}: we take linear combinations of edge modes that cancel the dominant asymptotic term(s), and then use the Gram-Schmidt process (with respect to the anticommutator) to construct anticommuting modes \cite{Verresen18}. We note that if the Fourier coefficients of the Wiener-Hopf factors themselves have a `nice' expansion, then singularity-filling will hold with no limiting $\nu_\star$ (see Appendix \ref{app:thm2}).

\textbf{Example.}
The long-range Kitaev chain corresponds to:
\begin{align}
    f_\textrm{LRK}(z)    &=\mu + J\left(\mathrm{Li}_\alpha(z)+\mathrm{Li}_\alpha(1/z)\right)\nonumber\\&+ \Delta\left(\mathrm{Li}_\beta(z)-\mathrm{Li}_\beta(1/z)\right).\label{eq:kitaev}
\end{align}
This model was studied for various choices of couplings in Refs.~\cite{Vodola15,Alecce17,Patrick17,Maity19,Jaeger20,Francica22}. 
Computing $(1/f(1/z))_n$ gives the asymptotic behaviour of the edge mode wavefunction in the $\omega=1$ case: $g_n = O( n^{-\Omega_0})$ for $\Omega_0=\min(\alpha,\beta)$, agreeing with results in the literature  (see Appendix \ref{app:Kitaev}).  
There are no other singularities, so Theorem \ref{thm2} implies that, for $0<\delta \omega <\lfloor\Omega_0\rfloor -2$, $f(z) = z^{\delta \omega} f_\textrm{LRK}(z)$ will have $\omega=1+\delta\omega$ edge modes with a basis decaying as $n^{-\Omega_0},n^{-(\Omega_0+1)},\dots,n^{-(\Omega_0+\delta\omega)}$.

\textbf{Singularity-filling for finite-size splitting.}
We now consider finite-size energy splittings for the edge modes. This quantity was considered in previous case studies of long-range Kitaev chains \cite{Vodola14,Jaeger20,Francica22}, but has not, to our knowledge, been explored in long-range systems with multiple edge modes (i.e., $|\omega|>1$).

In analogy with the singularity-filling for edge-mode wavefunctions above, we have a conjecture for the finite-size splittings for the edge modes. In this case the levels associated to singularities go up in steps of two. 
\begin{conjecture}[Splitting from singularity-filling]\label{conjecture}
Take an open chain $H_\textrm{BDI}$ of size $L$, where the related bulk Hamiltonian has winding number $\omega>0$ and $1/f(1/z)$ has singularities as defined above.

We conjecture that the $\omega$ finite-size edge modes have splittings $\eps_1 = \Theta( L^{-\nu_1}),\dots,\eps_\omega =\Theta(L^{-\nu_\omega})$ where the $\nu_k$ are the $\omega$ lowest levels $\mathcal{E}'_s(n)= {\Omega}_{k_s} +2n $ for $n\in \mathbb{Z}_{\geq 0}$. 

For $\omega<0$ analogous results hold where we replace $f(1/z)\rightarrow f(z)$.
 \end{conjecture}
This conjecture is based on numerical experiments (see Fig. \ref{fig:1}) and theoretical results (see below). The underlying theory indicates that for a family $f(z) = z^\omega f_0(z)$, there may exist an $\omega_\textrm{max}$ such that this holds only for $\omega<\omega_\textrm{max}$. In fact, given $\Omega_{\textrm{min}}>5$, and an assumption on the spectrum, we can prove the conjecture up to $\omega_\textrm{max}= 3$. However, empirically we expect the conjecture to hold more generally, as observed in Fig.~\ref{fig:1}.

The conjecture allows us to understand how finite-size effects hybridise the edge modes. For $\omega =1$ we see that the predicted splitting comes from the dominant singularity $\eps_1=\Theta( L^{-\Omega_\textrm{min}})$. Since this has the same asymptotics as the edge-mode wavefunction, this agrees with an intuitive connection between the spatial profile of the wavefunction and the induced splitting from the boundaries (see Appendix \ref{app:split}) that does not generically hold for the higher-winding case. For $\omega =2$ we expect to have two edge modes, one with $\eps_1=\Theta( L^{-\Omega_\textrm{min}})$ and one with either $\eps_2=\Theta( L^{-(\Omega_\textrm{min}+2)})$ or $\eps_2=\Theta( L^{-\Omega_{\textrm{next}}})$, depending on which has the slower decay. In the case of higher winding numbers, our conjecture predicts the hybridisation of the boundary modes, which is not in direct correspondence to the maximally localised basis identified in Theorem \ref{thm2}.

We can also make quantitative predictions without detailed calculation. Suppose we know for $\omega =1$ that we have an edge mode with splitting $\Theta(L^{-\nu})$, then for $\omega =2$ we infer that the second edge mode will have splitting $\Theta(L^{-\nu'})$ where $\nu \leq \nu'\leq \nu+2$. 
For $f(z) = z^n f_\textrm{LRK}(z)$ we have a singularity at $z=1$ only, and hence conjecture that splittings form a sequence 
$L^{-\Omega_0},L^{-(\Omega_0+2)}, \dots,L^{-(\Omega_0+2n)}$.

To justify the conjecture, consider models $f(z)= z^\omega f_0(z)$ with open boundary conditions; each such model has a corresponding single-particle (block Toeplitz) matrix, with determinant equal to $\prod_{j=1}^L (-\eps_j^2)$, where $\eps_j$ are single-particle energies. Assuming $(\alpha>1)$-decay, it can be shown, using Toeplitz determinants, that for the trivial model $f_0(z)$ this product is finite in the limit $L\rightarrow \infty$, while for $\omega \neq 0$, the corresponding determinant decays to zero with $L$ (with power depending on $\omega$ and Fourier coefficients of $1/f(z)$); see also Appendix \ref{app:filling}. Our method is to use the scaling of this determinant to predict the edge mode splitting. E.g., for $\omega =1$ we interpret:
\begin{align}
\prod_{j=1}^L (-\eps_j^2) = \textrm{const}\times L^{-\nu} (1+o(1)) 
\end{align}
as predicting a single edge mode with finite-size splitting $\eps_1  = \Theta(L^{-\nu}) $. For multiple edge modes (and $\omega>0$), we further assume inductively that the $\omega-1$ edge modes shared between the models $z^\omega f_0(z)$ and $z^{\omega-1} f_0(z)$ have the same energy splitting power-law in each model, and hence the additional decay in the determinant for $z^\omega f_0(z)$ comes from the $\omega$th edge mode \footnote{There is a symmetric claim in the case with $\omega<0$.}.

This is plausible since for periodic boundaries the models defined by $f(z)$ have spectrum independent of $\omega$, and we expect the system with open boundaries to differ from the bulk only `near the edge'. With finite-range interactions we believe this could be proved using results about eigenvalues of banded block Toeplitz matrices \cite{Ekstrom18}, for long-range chains we take it as an assumption that the scaling to zero with $L$ comes only from edge modes rather than the bulk band. In an earlier work the idea appeared in reverse: utilising the existence of exponentially-localised edge modes in short-range chains to predict asymptotics of block Toeplitz determinants \cite{Basor18}. 

We thus convert the question of finite-size edge mode splitting to a question about asymptotics of Toeplitz determinants. While there are several assumptions required to connect this theory to the edge mode splittings, the underlying singularity-filling picture for Toeplitz determinant asymptotics is in many cases fully rigorous. We outline some of these results in Appendix \ref{app:filling}; important references are \cite{Hartwig69,Widom90,Boettcher06, Boettcher06b}. 

\textbf{Novel topological probe.} A remarkable consequence is that the finite-size splitting of the lowest energy mode depends on the total number of edge modes. In fact, we can turn this into a probe of $\omega$: by perturbing a short-range chain $f_s(z)$ (with winding $\omega$) by a long-range test function, its finite-size splitting exponent will allow us to find $\omega$ (note that this is the scaling of the lowest one-particle energy, no further information about the spectrum is required). An example test function would be $ f_{\textrm{LRK}}(z)$, with $\Delta=0$. Then for the function $f(z)=f_s(z)+\epsilon f_{\textrm{LRK}}(z)$, for $\epsilon$ small, our picture gives a finite-size splitting $L^{-(\alpha+2(\lvert\omega\rvert-1))}$. 

\textbf{String-order parameters.}
We now consider the periodic chain. Define the finite fermion parity string by
$\mathcal{O}_0(n)=\prod_{m=1}^{n-1} \rmi\tilde\gamma_m\gamma_m $. Then consider further string operators, $\mathcal{O}_\kappa(n)$, of the form $ \mathcal{O}_0(n) \gamma_{n}\gamma_{n+1}\dots\gamma_{n+\kappa}$ for $\kappa>0$ and $\mathcal{O}_0(n) \tilde\gamma_{n}\dots\tilde\gamma_{n+\lvert\kappa\rvert-1}$
for $\kappa<0$ (up to phase factors).

It is know that the set of $\mathcal{O}_\kappa(n)$ form order parameters for the gapped phases in the short-range case \cite{Jones19}. In the long-range case we have:
\begin{theorem}[String order]\label{thm3}
Consider a gapped ($\alpha$>1)-decaying $H_{\textrm{BDI}}$, in the thermodynamic limit with periodic boundaries, and write $f(z)/|f(z)| = z^\omega \rme^{W(z)}$. Then:
\begin{align}
\lim_{N\rightarrow\infty} \lvert \langle \mathcal{O}_\kappa (1)\mathcal{O}_\kappa (N)\rangle \rvert =\delta_{\kappa\omega} \; \rme^{\sum_{k\geq0} k W_kW_{-k}}.
\end{align}
\end{theorem}
Thus the $\mathcal{O}_\kappa $ act as order parameters in the long-range case. The idea of the proof is as follows: the string-correlation functions $\langle \mathcal{O}_\kappa (1)\mathcal{O}_\kappa (N)\rangle$ are Toeplitz determinants generated by $z^{-\kappa} f(z)/\lvert f(z)\rvert$. The function $f(z)/\lvert f(z)\rvert$ generates the \emph{correlation matrix} of the chain, and it was proved in Ref.~\cite{Gong22} that for an $\alpha$-decaying chain with $\alpha>1$, the correlation matrix is $(\alpha-\eps)$-decaying for any $\eps>0$. This is sufficient regularity for us to use the results of Ref.~\cite{Hartwig69} to prove Theorem \ref{thm3} (see Appendix \ref{app:order}).

\textbf{Gap-closing and edge modes at critical points.}
For $H_{\textrm{BDI}}$ with finite-range couplings, topological edge modes can persist at critical points \cite{Verresen18,Verresen20}. We give some results in this direction for the long-range case.

Suppose we have a gapless bulk mode with dynamical critical exponent $z_{\textrm{dyn}}$. In the continuum limit, the dimension of the long-range term in the action $\delta S \sim \int \tilde\psi(x)\psi(y) (x-y)^{-\alpha}~\rmd t \rmd x \rmd y $ is $(z_{\textrm{dyn}}+1-\alpha)$, which is irrelevant for $\alpha>z_{\textrm{dyn}}+1$. On the lattice, we hence expect that for gapless models of the form $f_\textrm{crit}(z) = (z-1)^{z_{\textrm{dyn}}}f_\textrm{gap}(z)$ (which  has the aforementioned low-energy description if $f_\textrm{gap}(z)$ is non-vanishing on the unit circle), the edge modes will be stable as long as $f(z)$ is $(\alpha>z_{\textrm{dyn}}+1)$-decaying. Indeed, our Theorem \ref{thm1} can be adapted to show that this $f_\textrm{crit}(z)$ has $\omega$ localised edge modes where $\omega$ is the winding number of $f_\textrm{gap}(z)$. This follows from expanding $(z-1)^{z_{\textrm{dyn}}}$ in $f_\textrm{crit}(z)$, and interpreting this as a sum of $(z_\textrm{dyn}+1)$ gapped Hamiltonians, all sharing the same $\omega$ edge modes as per Theorem \ref{thm1}.

The above functional form can arise by interpolating between topologically distinct gapped Hamiltonians. For instance, between two phases with winding numbers $\omega=1$ and $\omega=2$, there will generically be a single gap-closing with a linearly-dispersing mode if $\alpha>2$. More precisely, if this occurs at momentum $k=0$, then $f_\textrm{gap}(z) := \frac{f(z)}{z-1}$ should define a gapped model with $\omega=1$. We can then apply the above discussion to infer the existence of the localised edge mode at criticality. We have confirmed this for an explicit example in Appendix \ref{app:gapless}.

\textbf{Outlook.} We have shown how general analytic methods can be used to establish the bulk-boundary correspondence in a class of long-range chains, and give insights into edge mode localisation and finite-size splitting. This included examples with $\alpha<1$ and certain gapless models.

Key questions remain within this class: what happens in the general case when $\alpha<1$ and the integer winding classification breaks down? Can we establish general stability results in critical lattice models, and do these coincide with our field-theoretic analysis? We expect extensions of analytic techniques used above to provide further insights. Moreover, it is worth exploring how broadly our results can be generalised, including to other free-fermion classes (beyond BDI and AIII) \cite{Schnyder08,Kitaev09,Ryu10} and higher-dimensional models.

The extension to long-range multi-band cases would be interesting, likely requiring block Toeplitz operators. In the short-range BDI and AIII classes, edge modes were constructed in Ref.~\cite{Balabanov21}, where the bulk topological index is the winding of the determinant of a chiral block of the Hamiltonian.

From the mathematical side, it would be most interesting to find a proof of the singularity-filling conjecture. It would be interesting to see if this picture generalises beyond the studied cases, perhaps even to interacting models with algebraically decaying edge modes, and whether their finite-size splitting also depends on the value of the topological invariant.

\section*{Acknowledgements}
We are grateful to Estelle Basor, Jon Keating and Ashvin Vishwanath for helpful correspondence and discussions, and to Dan Borgnia, Ruihua Fan and Rahul Sahay for useful discussions and comments on the manuscript. The work of NGJ was performed in part at the Aspen Center for Physics, which is supported by National Science Foundation grant PHY-1607611. The participation of NGJ at the Aspen Center for Physics was supported by a grant from the Simons Foundation. NGJ is grateful for support from the CMSA where this work was initiated and from KITP where this work was partially completed. RT is supported in part by the National Science Foundation under Grant No. NSF PHY-1748958. RV is supported by the Harvard Quantum Initiative Postdoctoral Fellowship in Science and Engineering and the Simons Collaboration on Ultra-Quantum Matter, which is a grant from the Simons Foundation (651440, Ashvin Vishwanath). 

\bibliography{arxiv.bbl}

\begin{thebibliography}{78}%
\makeatletter
\providecommand \@ifxundefined [1]{%
 \@ifx{#1\undefined}
}%
\providecommand \@ifnum [1]{%
 \ifnum #1\expandafter \@firstoftwo
 \else \expandafter \@secondoftwo
 \fi
}%
\providecommand \@ifx [1]{%
 \ifx #1\expandafter \@firstoftwo
 \else \expandafter \@secondoftwo
 \fi
}%
\providecommand \natexlab [1]{#1}%
\providecommand \enquote  [1]{``#1''}%
\providecommand \bibnamefont  [1]{#1}%
\providecommand \bibfnamefont [1]{#1}%
\providecommand \citenamefont [1]{#1}%
\providecommand \href@noop [0]{\@secondoftwo}%
\providecommand \href [0]{\begingroup \@sanitize@url \@href}%
\providecommand \@href[1]{\@@startlink{#1}\@@href}%
\providecommand \@@href[1]{\endgroup#1\@@endlink}%
\providecommand \@sanitize@url [0]{\catcode `\\12\catcode `\$12\catcode
  `\&12\catcode `\#12\catcode `\^12\catcode `\_12\catcode `\%12\relax}%
\providecommand \@@startlink[1]{}%
\providecommand \@@endlink[0]{}%
\providecommand \url  [0]{\begingroup\@sanitize@url \@url }%
\providecommand \@url [1]{\endgroup\@href {#1}{\urlprefix }}%
\providecommand \urlprefix  [0]{URL }%
\providecommand \Eprint [0]{\href }%
\providecommand \doibase [0]{https://doi.org/}%
\providecommand \selectlanguage [0]{\@gobble}%
\providecommand \bibinfo  [0]{\@secondoftwo}%
\providecommand \bibfield  [0]{\@secondoftwo}%
\providecommand \translation [1]{[#1]}%
\providecommand \BibitemOpen [0]{}%
\providecommand \bibitemStop [0]{}%
\providecommand \bibitemNoStop [0]{.\EOS\space}%
\providecommand \EOS [0]{\spacefactor3000\relax}%
\providecommand \BibitemShut  [1]{\csname bibitem#1\endcsname}%
\let\auto@bib@innerbib\@empty
\bibitem [{\citenamefont {Motrunich}\ \emph {et~al.}(2001)\citenamefont
  {Motrunich}, \citenamefont {Damle},\ and\ \citenamefont
  {Huse}}]{Motrunich01}%
  \BibitemOpen
  \bibfield  {author} {\bibinfo {author} {\bibfnamefont {O.}~\bibnamefont
  {Motrunich}}, \bibinfo {author} {\bibfnamefont {K.}~\bibnamefont {Damle}},\
  and\ \bibinfo {author} {\bibfnamefont {D.~A.}\ \bibnamefont {Huse}},\
  }\bibfield  {title} {\bibinfo {title} {Griffiths effects and quantum critical
  points in dirty superconductors without spin-rotation invariance:
  One-dimensional examples},\ }\href
  {https://doi.org/10.1103/PhysRevB.63.224204} {\bibfield  {journal} {\bibinfo
  {journal} {Phys. Rev. B}\ }\textbf {\bibinfo {volume} {63}},\ \bibinfo
  {pages} {224204} (\bibinfo {year} {2001})}\BibitemShut {NoStop}%
\bibitem [{\citenamefont {Ryu}\ and\ \citenamefont {Hatsugai}(2002)}]{Ryu02}%
  \BibitemOpen
  \bibfield  {author} {\bibinfo {author} {\bibfnamefont {S.}~\bibnamefont
  {Ryu}}\ and\ \bibinfo {author} {\bibfnamefont {Y.}~\bibnamefont {Hatsugai}},\
  }\bibfield  {title} {\bibinfo {title} {Topological origin of zero-energy edge
  states in particle-hole symmetric systems},\ }\href
  {https://doi.org/10.1103/PhysRevLett.89.077002} {\bibfield  {journal}
  {\bibinfo  {journal} {Phys. Rev. Lett.}\ }\textbf {\bibinfo {volume} {89}},\
  \bibinfo {pages} {077002} (\bibinfo {year} {2002})}\BibitemShut {NoStop}%
\bibitem [{\citenamefont {Li}\ and\ \citenamefont {Haldane}(2008)}]{Li08}%
  \BibitemOpen
  \bibfield  {author} {\bibinfo {author} {\bibfnamefont {H.}~\bibnamefont
  {Li}}\ and\ \bibinfo {author} {\bibfnamefont {F.~D.~M.}\ \bibnamefont
  {Haldane}},\ }\bibfield  {title} {\bibinfo {title} {{Entanglement Spectrum as
  a Generalization of Entanglement Entropy: Identification of Topological Order
  in Non-Abelian Fractional Quantum Hall Effect States}},\ }\href
  {https://doi.org/10.1103/PhysRevLett.101.010504} {\bibfield  {journal}
  {\bibinfo  {journal} {Phys. Rev. Lett.}\ }\textbf {\bibinfo {volume} {101}},\
  \bibinfo {pages} {010504} (\bibinfo {year} {2008})}\BibitemShut {NoStop}%
\bibitem [{\citenamefont {Schnyder}\ \emph {et~al.}(2008)\citenamefont
  {Schnyder}, \citenamefont {Ryu}, \citenamefont {Furusaki},\ and\
  \citenamefont {Ludwig}}]{Schnyder08}%
  \BibitemOpen
  \bibfield  {author} {\bibinfo {author} {\bibfnamefont {A.~P.}\ \bibnamefont
  {Schnyder}}, \bibinfo {author} {\bibfnamefont {S.}~\bibnamefont {Ryu}},
  \bibinfo {author} {\bibfnamefont {A.}~\bibnamefont {Furusaki}},\ and\
  \bibinfo {author} {\bibfnamefont {A.~W.~W.}\ \bibnamefont {Ludwig}},\
  }\bibfield  {title} {\bibinfo {title} {Classification of topological
  insulators and superconductors in three spatial dimensions},\ }\href
  {https://doi.org/10.1103/PhysRevB.78.195125} {\bibfield  {journal} {\bibinfo
  {journal} {Phys. Rev. B}\ }\textbf {\bibinfo {volume} {78}},\ \bibinfo
  {pages} {195125} (\bibinfo {year} {2008})}\BibitemShut {NoStop}%
\bibitem [{\citenamefont {Schnyder}\ \emph {et~al.}(2009)\citenamefont
  {Schnyder}, \citenamefont {Ryu}, \citenamefont {Furusaki},\ and\
  \citenamefont {Ludwig}}]{Schnyder09}%
  \BibitemOpen
  \bibfield  {author} {\bibinfo {author} {\bibfnamefont {A.~P.}\ \bibnamefont
  {Schnyder}}, \bibinfo {author} {\bibfnamefont {S.}~\bibnamefont {Ryu}},
  \bibinfo {author} {\bibfnamefont {A.}~\bibnamefont {Furusaki}},\ and\
  \bibinfo {author} {\bibfnamefont {A.~W.~W.}\ \bibnamefont {Ludwig}},\
  }\bibfield  {title} {\bibinfo {title} {Classification of topological
  insulators and superconductors},\ }\href {https://doi.org/10.1063/1.3149481}
  {\bibfield  {journal} {\bibinfo  {journal} {AIP Conference Proceedings}\
  }\textbf {\bibinfo {volume} {1134}},\ \bibinfo {pages} {10} (\bibinfo {year}
  {2009})},\ \Eprint
  {https://arxiv.org/abs/https://aip.scitation.org/doi/pdf/10.1063/1.3149481}
  {https://aip.scitation.org/doi/pdf/10.1063/1.3149481} \BibitemShut {NoStop}%
\bibitem [{\citenamefont {{Hatsugai}}(2009)}]{Hatsugai09}%
  \BibitemOpen
  \bibfield  {author} {\bibinfo {author} {\bibfnamefont {Y.}~\bibnamefont
  {{Hatsugai}}},\ }\bibfield  {title} {\bibinfo {title} {{Bulk-edge
  correspondence in graphene with/without magnetic field: Chiral symmetry,
  Dirac fermions and edge states}},\ }\href
  {https://doi.org/10.1016/j.ssc.2009.02.055} {\bibfield  {journal} {\bibinfo
  {journal} {Solid State Communications}\ }\textbf {\bibinfo {volume} {149}},\
  \bibinfo {pages} {1061} (\bibinfo {year} {2009})}\BibitemShut {NoStop}%
\bibitem [{\citenamefont {Hasan}\ and\ \citenamefont {Kane}(2010)}]{Hasan10}%
  \BibitemOpen
  \bibfield  {author} {\bibinfo {author} {\bibfnamefont {M.~Z.}\ \bibnamefont
  {Hasan}}\ and\ \bibinfo {author} {\bibfnamefont {C.~L.}\ \bibnamefont
  {Kane}},\ }\bibfield  {title} {\bibinfo {title} {Colloquium: Topological
  insulators},\ }\href {https://doi.org/10.1103/RevModPhys.82.3045} {\bibfield
  {journal} {\bibinfo  {journal} {Rev. Mod. Phys.}\ }\textbf {\bibinfo {volume}
  {82}},\ \bibinfo {pages} {3045} (\bibinfo {year} {2010})}\BibitemShut
  {NoStop}%
\bibitem [{\citenamefont {Ryu}\ \emph {et~al.}(2010)\citenamefont {Ryu},
  \citenamefont {Schnyder}, \citenamefont {Furusaki},\ and\ \citenamefont
  {Ludwig}}]{Ryu10}%
  \BibitemOpen
  \bibfield  {author} {\bibinfo {author} {\bibfnamefont {S.}~\bibnamefont
  {Ryu}}, \bibinfo {author} {\bibfnamefont {A.~P.}\ \bibnamefont {Schnyder}},
  \bibinfo {author} {\bibfnamefont {A.}~\bibnamefont {Furusaki}},\ and\
  \bibinfo {author} {\bibfnamefont {A.~W.~W.}\ \bibnamefont {Ludwig}},\
  }\bibfield  {title} {\bibinfo {title} {Topological insulators and
  superconductors: tenfold way and dimensional hierarchy},\ }\href
  {https://doi.org/10.1088/1367-2630/12/6/065010} {\bibfield  {journal}
  {\bibinfo  {journal} {New Journal of Physics}\ }\textbf {\bibinfo {volume}
  {12}},\ \bibinfo {pages} {065010} (\bibinfo {year} {2010})}\BibitemShut
  {NoStop}%
\bibitem [{\citenamefont {{Delplace}}\ \emph {et~al.}(2011)\citenamefont
  {{Delplace}}, \citenamefont {{Ullmo}},\ and\ \citenamefont
  {{Montambaux}}}]{Delplace11}%
  \BibitemOpen
  \bibfield  {author} {\bibinfo {author} {\bibfnamefont {P.}~\bibnamefont
  {{Delplace}}}, \bibinfo {author} {\bibfnamefont {D.}~\bibnamefont
  {{Ullmo}}},\ and\ \bibinfo {author} {\bibfnamefont {G.}~\bibnamefont
  {{Montambaux}}},\ }\bibfield  {title} {\bibinfo {title} {{Zak phase and the
  existence of edge states in graphene}},\ }\href
  {https://doi.org/10.1103/PhysRevB.84.195452} {\bibfield  {journal} {\bibinfo
  {journal} {\prb}\ }\textbf {\bibinfo {volume} {84}},\ \bibinfo {eid} {195452}
  (\bibinfo {year} {2011})}\BibitemShut {NoStop}%
\bibitem [{\citenamefont {Mong}\ and\ \citenamefont
  {Shivamoggi}(2011)}]{Mong11}%
  \BibitemOpen
  \bibfield  {author} {\bibinfo {author} {\bibfnamefont {R.~S.~K.}\
  \bibnamefont {Mong}}\ and\ \bibinfo {author} {\bibfnamefont {V.}~\bibnamefont
  {Shivamoggi}},\ }\bibfield  {title} {\bibinfo {title} {{Edge states and the
  bulk-boundary correspondence in Dirac Hamiltonians}},\ }\href
  {https://doi.org/10.1103/PhysRevB.83.125109} {\bibfield  {journal} {\bibinfo
  {journal} {Phys. Rev. B}\ }\textbf {\bibinfo {volume} {83}},\ \bibinfo
  {pages} {125109} (\bibinfo {year} {2011})}\BibitemShut {NoStop}%
\bibitem [{\citenamefont {{Tanaka}}\ \emph {et~al.}(2012)\citenamefont
  {{Tanaka}}, \citenamefont {{Sato}},\ and\ \citenamefont
  {{Nagaosa}}}]{Tanaka12}%
  \BibitemOpen
  \bibfield  {author} {\bibinfo {author} {\bibfnamefont {Y.}~\bibnamefont
  {{Tanaka}}}, \bibinfo {author} {\bibfnamefont {M.}~\bibnamefont {{Sato}}},\
  and\ \bibinfo {author} {\bibfnamefont {N.}~\bibnamefont {{Nagaosa}}},\
  }\bibfield  {title} {\bibinfo {title} {{Symmetry and Topology in
  Superconductors ---Odd-Frequency Pairing and Edge States---}},\ }\href
  {https://doi.org/10.1143/JPSJ.81.011013} {\bibfield  {journal} {\bibinfo
  {journal} {Journal of the Physical Society of Japan}\ }\textbf {\bibinfo
  {volume} {81}},\ \bibinfo {pages} {011013} (\bibinfo {year}
  {2012})}\BibitemShut {NoStop}%
\bibitem [{\citenamefont {{Graf}}\ and\ \citenamefont
  {{Porta}}(2013)}]{Graf13}%
  \BibitemOpen
  \bibfield  {author} {\bibinfo {author} {\bibfnamefont {G.~M.}\ \bibnamefont
  {{Graf}}}\ and\ \bibinfo {author} {\bibfnamefont {M.}~\bibnamefont
  {{Porta}}},\ }\bibfield  {title} {\bibinfo {title} {{Bulk-Edge Correspondence
  for Two-Dimensional Topological Insulators}},\ }\href
  {https://doi.org/10.1007/s00220-013-1819-6} {\bibfield  {journal} {\bibinfo
  {journal} {Communications in Mathematical Physics}\ }\textbf {\bibinfo
  {volume} {324}},\ \bibinfo {pages} {851} (\bibinfo {year}
  {2013})}\BibitemShut {NoStop}%
\bibitem [{\citenamefont {Bernevig}\ and\ \citenamefont
  {Neupert}(2017)}]{Bernevig15}%
  \BibitemOpen
  \bibfield  {author} {\bibinfo {author} {\bibfnamefont {A.}~\bibnamefont
  {Bernevig}}\ and\ \bibinfo {author} {\bibfnamefont {T.}~\bibnamefont
  {Neupert}},\ }\bibfield  {title} {\bibinfo {title} {{Topological
  Superconductors and Category Theory}},\ }in\ \href@noop {} {\emph {\bibinfo
  {booktitle} {{Lecture Notes of the Les Houches Summer School: Topological
  Aspects of Condensed Matter Physics}}}}\ (\bibinfo {year} {2017})\ pp.\
  \bibinfo {pages} {63--121}\BibitemShut {NoStop}%
\bibitem [{\citenamefont {Asb{\'{o}}th}\ \emph {et~al.}(2016)\citenamefont
  {Asb{\'{o}}th}, \citenamefont {Oroszl{\'{a}}ny},\ and\ \citenamefont
  {P{\'{a}}lyi}}]{Asboth16}%
  \BibitemOpen
  \bibfield  {author} {\bibinfo {author} {\bibfnamefont {J.~K.}\ \bibnamefont
  {Asb{\'{o}}th}}, \bibinfo {author} {\bibfnamefont {L.}~\bibnamefont
  {Oroszl{\'{a}}ny}},\ and\ \bibinfo {author} {\bibfnamefont {A.}~\bibnamefont
  {P{\'{a}}lyi}},\ }\href {https://doi.org/10.1007/978-3-319-25607-8} {\emph
  {\bibinfo {title} {A Short Course on Topological Insulators}}}\ (\bibinfo
  {publisher} {Springer International Publishing},\ \bibinfo {year}
  {2016})\BibitemShut {NoStop}%
\bibitem [{\citenamefont {{Peng}}\ \emph {et~al.}(2017)\citenamefont {{Peng}},
  \citenamefont {{Bao}},\ and\ \citenamefont {{von Oppen}}}]{Peng17}%
  \BibitemOpen
  \bibfield  {author} {\bibinfo {author} {\bibfnamefont {Y.}~\bibnamefont
  {{Peng}}}, \bibinfo {author} {\bibfnamefont {Y.}~\bibnamefont {{Bao}}},\ and\
  \bibinfo {author} {\bibfnamefont {F.}~\bibnamefont {{von Oppen}}},\
  }\bibfield  {title} {\bibinfo {title} {{Boundary Green functions of
  topological insulators and superconductors}},\ }\href
  {https://doi.org/10.1103/PhysRevB.95.235143} {\bibfield  {journal} {\bibinfo
  {journal} {\prb}\ }\textbf {\bibinfo {volume} {95}},\ \bibinfo {eid} {235143}
  (\bibinfo {year} {2017})}\BibitemShut {NoStop}%
\bibitem [{\citenamefont {{Sedlmayr}}\ \emph {et~al.}(2017)\citenamefont
  {{Sedlmayr}}, \citenamefont {{Kaladzhyan}}, \citenamefont {{Dutreix}},\ and\
  \citenamefont {{Bena}}}]{Sedlmayr17}%
  \BibitemOpen
  \bibfield  {author} {\bibinfo {author} {\bibfnamefont {N.}~\bibnamefont
  {{Sedlmayr}}}, \bibinfo {author} {\bibfnamefont {V.}~\bibnamefont
  {{Kaladzhyan}}}, \bibinfo {author} {\bibfnamefont {C.}~\bibnamefont
  {{Dutreix}}},\ and\ \bibinfo {author} {\bibfnamefont {C.}~\bibnamefont
  {{Bena}}},\ }\bibfield  {title} {\bibinfo {title} {{Bulk boundary
  correspondence and the existence of Majorana bound states on the edges of 2D
  topological superconductors}},\ }\href
  {https://doi.org/10.1103/PhysRevB.96.184516} {\bibfield  {journal} {\bibinfo
  {journal} {\prb}\ }\textbf {\bibinfo {volume} {96}},\ \bibinfo {eid} {184516}
  (\bibinfo {year} {2017})}\BibitemShut {NoStop}%
\bibitem [{\citenamefont {{Rhim}}\ \emph {et~al.}(2018)\citenamefont {{Rhim}},
  \citenamefont {{Bardarson}},\ and\ \citenamefont {{Slager}}}]{Rhim18}%
  \BibitemOpen
  \bibfield  {author} {\bibinfo {author} {\bibfnamefont {J.-W.}\ \bibnamefont
  {{Rhim}}}, \bibinfo {author} {\bibfnamefont {J.~H.}\ \bibnamefont
  {{Bardarson}}},\ and\ \bibinfo {author} {\bibfnamefont {R.-J.}\ \bibnamefont
  {{Slager}}},\ }\bibfield  {title} {\bibinfo {title} {{Unified bulk-boundary
  correspondence for band insulators}},\ }\href
  {https://doi.org/10.1103/PhysRevB.97.115143} {\bibfield  {journal} {\bibinfo
  {journal} {\prb}\ }\textbf {\bibinfo {volume} {97}},\ \bibinfo {eid} {115143}
  (\bibinfo {year} {2018})}\BibitemShut {NoStop}%
\bibitem [{\citenamefont {Kitaev}(2009)}]{Kitaev09}%
  \BibitemOpen
  \bibfield  {author} {\bibinfo {author} {\bibfnamefont {A.}~\bibnamefont
  {Kitaev}},\ }\bibfield  {title} {\bibinfo {title} {Periodic table for
  topological insulators and superconductors},\ }in\ \href
  {https://doi.org/10.1063/1.3149495} {\emph {\bibinfo {booktitle} {{AIP}
  Conference Proceedings}}}\ (\bibinfo  {publisher} {{AIP}},\ \bibinfo {year}
  {2009})\BibitemShut {NoStop}%
\bibitem [{\citenamefont {Fidkowski}\ and\ \citenamefont
  {Kitaev}(2010)}]{Fidkowski10}%
  \BibitemOpen
  \bibfield  {author} {\bibinfo {author} {\bibfnamefont {L.}~\bibnamefont
  {Fidkowski}}\ and\ \bibinfo {author} {\bibfnamefont {A.}~\bibnamefont
  {Kitaev}},\ }\bibfield  {title} {\bibinfo {title} {Effects of interactions on
  the topological classification of free fermion systems},\ }\bibfield
  {journal} {\bibinfo  {journal} {Physical Review B}\ }\textbf {\bibinfo
  {volume} {81}},\ \href {https://doi.org/10.1103/physrevb.81.134509}
  {10.1103/physrevb.81.134509} (\bibinfo {year} {2010})\BibitemShut {NoStop}%
\bibitem [{\citenamefont {DeGottardi}\ \emph {et~al.}(2013)\citenamefont
  {DeGottardi}, \citenamefont {Thakurathi}, \citenamefont {Vishveshwara},\ and\
  \citenamefont {Sen}}]{DeGottardi13}%
  \BibitemOpen
  \bibfield  {author} {\bibinfo {author} {\bibfnamefont {W.}~\bibnamefont
  {DeGottardi}}, \bibinfo {author} {\bibfnamefont {M.}~\bibnamefont
  {Thakurathi}}, \bibinfo {author} {\bibfnamefont {S.}~\bibnamefont
  {Vishveshwara}},\ and\ \bibinfo {author} {\bibfnamefont {D.}~\bibnamefont
  {Sen}},\ }\bibfield  {title} {\bibinfo {title} {Majorana fermions in
  superconducting wires: Effects of long-range hopping, broken time-reversal
  symmetry, and potential landscapes},\ }\bibfield  {journal} {\bibinfo
  {journal} {Physical Review B}\ }\textbf {\bibinfo {volume} {88}},\ \href
  {https://doi.org/10.1103/physrevb.88.165111} {10.1103/physrevb.88.165111}
  (\bibinfo {year} {2013})\BibitemShut {NoStop}%
\bibitem [{\citenamefont {Verresen}\ \emph {et~al.}(2018)\citenamefont
  {Verresen}, \citenamefont {Jones},\ and\ \citenamefont
  {Pollmann}}]{Verresen18}%
  \BibitemOpen
  \bibfield  {author} {\bibinfo {author} {\bibfnamefont {R.}~\bibnamefont
  {Verresen}}, \bibinfo {author} {\bibfnamefont {N.~G.}\ \bibnamefont
  {Jones}},\ and\ \bibinfo {author} {\bibfnamefont {F.}~\bibnamefont
  {Pollmann}},\ }\bibfield  {title} {\bibinfo {title} {{Topology and Edge Modes
  in Quantum Critical Chains}},\ }\bibfield  {journal} {\bibinfo  {journal}
  {Physical Review Letters}\ }\textbf {\bibinfo {volume} {120}},\ \href
  {https://doi.org/10.1103/physrevlett.120.057001}
  {10.1103/physrevlett.120.057001} (\bibinfo {year} {2018})\BibitemShut
  {NoStop}%
\bibitem [{\citenamefont {Maity}\ \emph {et~al.}(2019)\citenamefont {Maity},
  \citenamefont {Bhattacharya},\ and\ \citenamefont {Dutta}}]{Maity19}%
  \BibitemOpen
  \bibfield  {author} {\bibinfo {author} {\bibfnamefont {S.}~\bibnamefont
  {Maity}}, \bibinfo {author} {\bibfnamefont {U.}~\bibnamefont
  {Bhattacharya}},\ and\ \bibinfo {author} {\bibfnamefont {A.}~\bibnamefont
  {Dutta}},\ }\bibfield  {title} {\bibinfo {title} {One-dimensional quantum
  many body systems with long-range interactions},\ }\href
  {https://doi.org/10.1088/1751-8121/ab5634} {\bibfield  {journal} {\bibinfo
  {journal} {Journal of Physics A: Mathematical and Theoretical}\ }\textbf
  {\bibinfo {volume} {53}},\ \bibinfo {pages} {013001} (\bibinfo {year}
  {2019})}\BibitemShut {NoStop}%
\bibitem [{\citenamefont {Defenu}\ \emph {et~al.}(2021)\citenamefont {Defenu},
  \citenamefont {Donner}, \citenamefont {Macrì}, \citenamefont {Pagano},
  \citenamefont {Ruffo},\ and\ \citenamefont {Trombettoni}}]{Defenu21}%
  \BibitemOpen
  \bibfield  {author} {\bibinfo {author} {\bibfnamefont {N.}~\bibnamefont
  {Defenu}}, \bibinfo {author} {\bibfnamefont {T.}~\bibnamefont {Donner}},
  \bibinfo {author} {\bibfnamefont {T.}~\bibnamefont {Macrì}}, \bibinfo
  {author} {\bibfnamefont {G.}~\bibnamefont {Pagano}}, \bibinfo {author}
  {\bibfnamefont {S.}~\bibnamefont {Ruffo}},\ and\ \bibinfo {author}
  {\bibfnamefont {A.}~\bibnamefont {Trombettoni}},\ }\href
  {https://doi.org/10.48550/ARXIV.2109.01063} {\bibinfo {title} {Long-range
  interacting quantum systems}} (\bibinfo {year} {2021})\BibitemShut {NoStop}%
\bibitem [{\citenamefont {Pientka}\ \emph {et~al.}(2013)\citenamefont
  {Pientka}, \citenamefont {Glazman},\ and\ \citenamefont {von
  Oppen}}]{Pientka13}%
  \BibitemOpen
  \bibfield  {author} {\bibinfo {author} {\bibfnamefont {F.}~\bibnamefont
  {Pientka}}, \bibinfo {author} {\bibfnamefont {L.~I.}\ \bibnamefont
  {Glazman}},\ and\ \bibinfo {author} {\bibfnamefont {F.}~\bibnamefont {von
  Oppen}},\ }\bibfield  {title} {\bibinfo {title} {{Topological superconducting
  phase in helical Shiba chains}},\ }\href
  {https://doi.org/10.1103/PhysRevB.88.155420} {\bibfield  {journal} {\bibinfo
  {journal} {Phys. Rev. B}\ }\textbf {\bibinfo {volume} {88}},\ \bibinfo
  {pages} {155420} (\bibinfo {year} {2013})}\BibitemShut {NoStop}%
\bibitem [{\citenamefont {Gong}\ \emph {et~al.}(2017)\citenamefont {Gong},
  \citenamefont {Foss-Feig}, \citenamefont {Brand\~ao},\ and\ \citenamefont
  {Gorshkov}}]{Gong17}%
  \BibitemOpen
  \bibfield  {author} {\bibinfo {author} {\bibfnamefont {Z.-X.}\ \bibnamefont
  {Gong}}, \bibinfo {author} {\bibfnamefont {M.}~\bibnamefont {Foss-Feig}},
  \bibinfo {author} {\bibfnamefont {F.~G. S.~L.}\ \bibnamefont {Brand\~ao}},\
  and\ \bibinfo {author} {\bibfnamefont {A.~V.}\ \bibnamefont {Gorshkov}},\
  }\bibfield  {title} {\bibinfo {title} {Entanglement area laws for long-range
  interacting systems},\ }\href
  {https://doi.org/10.1103/PhysRevLett.119.050501} {\bibfield  {journal}
  {\bibinfo  {journal} {Phys. Rev. Lett.}\ }\textbf {\bibinfo {volume} {119}},\
  \bibinfo {pages} {050501} (\bibinfo {year} {2017})}\BibitemShut {NoStop}%
\bibitem [{\citenamefont {Lepori}\ \emph {et~al.}(2016)\citenamefont {Lepori},
  \citenamefont {Vodola}, \citenamefont {Pupillo}, \citenamefont {Gori},\ and\
  \citenamefont {Trombettoni}}]{Lepori16}%
  \BibitemOpen
  \bibfield  {author} {\bibinfo {author} {\bibfnamefont {L.}~\bibnamefont
  {Lepori}}, \bibinfo {author} {\bibfnamefont {D.}~\bibnamefont {Vodola}},
  \bibinfo {author} {\bibfnamefont {G.}~\bibnamefont {Pupillo}}, \bibinfo
  {author} {\bibfnamefont {G.}~\bibnamefont {Gori}},\ and\ \bibinfo {author}
  {\bibfnamefont {A.}~\bibnamefont {Trombettoni}},\ }\bibfield  {title}
  {\bibinfo {title} {Effective theory and breakdown of conformal symmetry in a
  long-range quantum chain},\ }\href
  {https://doi.org/10.1016/j.aop.2016.07.026} {\bibfield  {journal} {\bibinfo
  {journal} {Annals of Physics}\ }\textbf {\bibinfo {volume} {374}},\ \bibinfo
  {pages} {35} (\bibinfo {year} {2016})}\BibitemShut {NoStop}%
\bibitem [{\citenamefont {Kitaev}(2001)}]{Kitaev01}%
  \BibitemOpen
  \bibfield  {author} {\bibinfo {author} {\bibfnamefont {A.}~\bibnamefont
  {Kitaev}},\ }\bibfield  {title} {\bibinfo {title} {{Unpaired Majorana
  fermions in quantum wires}},\ }\href@noop {} {\bibfield  {journal} {\bibinfo
  {journal} {Physics-Uspekhi}\ }\textbf {\bibinfo {volume} {44}},\ \bibinfo
  {pages} {131} (\bibinfo {year} {2001})}\BibitemShut {NoStop}%
\bibitem [{\citenamefont {Vodola}\ \emph {et~al.}(2014)\citenamefont {Vodola},
  \citenamefont {Lepori}, \citenamefont {Ercolessi}, \citenamefont {Gorshkov},\
  and\ \citenamefont {Pupillo}}]{Vodola14}%
  \BibitemOpen
  \bibfield  {author} {\bibinfo {author} {\bibfnamefont {D.}~\bibnamefont
  {Vodola}}, \bibinfo {author} {\bibfnamefont {L.}~\bibnamefont {Lepori}},
  \bibinfo {author} {\bibfnamefont {E.}~\bibnamefont {Ercolessi}}, \bibinfo
  {author} {\bibfnamefont {A.~V.}\ \bibnamefont {Gorshkov}},\ and\ \bibinfo
  {author} {\bibfnamefont {G.}~\bibnamefont {Pupillo}},\ }\bibfield  {title}
  {\bibinfo {title} {{Kitaev Chains with Long-Range Pairing}},\ }\bibfield
  {journal} {\bibinfo  {journal} {Physical Review Letters}\ }\textbf {\bibinfo
  {volume} {113}},\ \href {https://doi.org/10.1103/physrevlett.113.156402}
  {10.1103/physrevlett.113.156402} (\bibinfo {year} {2014})\BibitemShut
  {NoStop}%
\bibitem [{\citenamefont {Vodola}\ \emph {et~al.}(2015)\citenamefont {Vodola},
  \citenamefont {Lepori}, \citenamefont {Ercolessi},\ and\ \citenamefont
  {Pupillo}}]{Vodola15}%
  \BibitemOpen
  \bibfield  {author} {\bibinfo {author} {\bibfnamefont {D.}~\bibnamefont
  {Vodola}}, \bibinfo {author} {\bibfnamefont {L.}~\bibnamefont {Lepori}},
  \bibinfo {author} {\bibfnamefont {E.}~\bibnamefont {Ercolessi}},\ and\
  \bibinfo {author} {\bibfnamefont {G.}~\bibnamefont {Pupillo}},\ }\bibfield
  {title} {\bibinfo {title} {{Long-range Ising and Kitaev models: phases,
  correlations and edge modes}},\ }\href
  {https://doi.org/10.1088/1367-2630/18/1/015001} {\bibfield  {journal}
  {\bibinfo  {journal} {New Journal of Physics}\ }\textbf {\bibinfo {volume}
  {18}},\ \bibinfo {pages} {015001} (\bibinfo {year} {2015})}\BibitemShut
  {NoStop}%
\bibitem [{\citenamefont {Lepori}\ and\ \citenamefont
  {Dell’Anna}(2017)}]{Lepori17}%
  \BibitemOpen
  \bibfield  {author} {\bibinfo {author} {\bibfnamefont {L.}~\bibnamefont
  {Lepori}}\ and\ \bibinfo {author} {\bibfnamefont {L.}~\bibnamefont
  {Dell’Anna}},\ }\bibfield  {title} {\bibinfo {title} {Long-range
  topological insulators and weakened bulk-boundary correspondence},\ }\href
  {https://doi.org/10.1088/1367-2630/aa84d0} {\bibfield  {journal} {\bibinfo
  {journal} {New Journal of Physics}\ }\textbf {\bibinfo {volume} {19}},\
  \bibinfo {pages} {103030} (\bibinfo {year} {2017})}\BibitemShut {NoStop}%
\bibitem [{\citenamefont {Alecce}\ and\ \citenamefont
  {Dell{\textquotesingle}Anna}(2017)}]{Alecce17}%
  \BibitemOpen
  \bibfield  {author} {\bibinfo {author} {\bibfnamefont {A.}~\bibnamefont
  {Alecce}}\ and\ \bibinfo {author} {\bibfnamefont {L.}~\bibnamefont
  {Dell{\textquotesingle}Anna}},\ }\bibfield  {title} {\bibinfo {title}
  {{Extended Kitaev chain with longer-range hopping and pairing}},\ }\bibfield
  {journal} {\bibinfo  {journal} {Physical Review B}\ }\textbf {\bibinfo
  {volume} {95}},\ \href {https://doi.org/10.1103/physrevb.95.195160}
  {10.1103/physrevb.95.195160} (\bibinfo {year} {2017})\BibitemShut {NoStop}%
\bibitem [{\citenamefont {Patrick}\ \emph {et~al.}(2017)\citenamefont
  {Patrick}, \citenamefont {Neupert},\ and\ \citenamefont
  {Pachos}}]{Patrick17}%
  \BibitemOpen
  \bibfield  {author} {\bibinfo {author} {\bibfnamefont {K.}~\bibnamefont
  {Patrick}}, \bibinfo {author} {\bibfnamefont {T.}~\bibnamefont {Neupert}},\
  and\ \bibinfo {author} {\bibfnamefont {J.~K.}\ \bibnamefont {Pachos}},\
  }\bibfield  {title} {\bibinfo {title} {Topological quantum liquids with
  long-range couplings},\ }\bibfield  {journal} {\bibinfo  {journal} {Physical
  Review Letters}\ }\textbf {\bibinfo {volume} {118}},\ \href
  {https://doi.org/10.1103/physrevlett.118.267002}
  {10.1103/physrevlett.118.267002} (\bibinfo {year} {2017})\BibitemShut
  {NoStop}%
\bibitem [{\citenamefont {Jäger}\ \emph {et~al.}(2020)\citenamefont {Jäger},
  \citenamefont {Dell{\textquotesingle}Anna},\ and\ \citenamefont
  {Morigi}}]{Jaeger20}%
  \BibitemOpen
  \bibfield  {author} {\bibinfo {author} {\bibfnamefont {S.~B.}\ \bibnamefont
  {Jäger}}, \bibinfo {author} {\bibfnamefont {L.}~\bibnamefont
  {Dell{\textquotesingle}Anna}},\ and\ \bibinfo {author} {\bibfnamefont
  {G.}~\bibnamefont {Morigi}},\ }\bibfield  {title} {\bibinfo {title} {{Edge
  states of the long-range Kitaev chain: An analytical study}},\ }\bibfield
  {journal} {\bibinfo  {journal} {Physical Review B}\ }\textbf {\bibinfo
  {volume} {102}},\ \href {https://doi.org/10.1103/physrevb.102.035152}
  {10.1103/physrevb.102.035152} (\bibinfo {year} {2020})\BibitemShut {NoStop}%
\bibitem [{\citenamefont {Kartik}\ \emph {et~al.}(2021)\citenamefont {Kartik},
  \citenamefont {Kumar}, \citenamefont {Rahul}, \citenamefont {Roy},\ and\
  \citenamefont {Sarkar}}]{Kartik21}%
  \BibitemOpen
  \bibfield  {author} {\bibinfo {author} {\bibfnamefont {Y.~R.}\ \bibnamefont
  {Kartik}}, \bibinfo {author} {\bibfnamefont {R.~R.}\ \bibnamefont {Kumar}},
  \bibinfo {author} {\bibfnamefont {S.}~\bibnamefont {Rahul}}, \bibinfo
  {author} {\bibfnamefont {N.}~\bibnamefont {Roy}},\ and\ \bibinfo {author}
  {\bibfnamefont {S.}~\bibnamefont {Sarkar}},\ }\bibfield  {title} {\bibinfo
  {title} {{Topological quantum phase transitions and criticality in a
  longer-range Kitaev chain}},\ }\href
  {https://doi.org/10.1103/PhysRevB.104.075113} {\bibfield  {journal} {\bibinfo
   {journal} {Phys. Rev. B}\ }\textbf {\bibinfo {volume} {104}},\ \bibinfo
  {pages} {075113} (\bibinfo {year} {2021})}\BibitemShut {NoStop}%
\bibitem [{\citenamefont {Sadhukhan}\ and\ \citenamefont
  {Dziarmaga}(2021)}]{Sadhukhan21}%
  \BibitemOpen
  \bibfield  {author} {\bibinfo {author} {\bibfnamefont {D.}~\bibnamefont
  {Sadhukhan}}\ and\ \bibinfo {author} {\bibfnamefont {J.}~\bibnamefont
  {Dziarmaga}},\ }\href {https://doi.org/10.48550/ARXIV.2107.02508} {\bibinfo
  {title} {Is there a correlation length in a model with long-range
  interactions?}} (\bibinfo {year} {2021})\BibitemShut {NoStop}%
\bibitem [{\citenamefont {Francica}\ and\ \citenamefont
  {Dell{\textquotesingle}Anna}(2022)}]{Francica22}%
  \BibitemOpen
  \bibfield  {author} {\bibinfo {author} {\bibfnamefont {G.}~\bibnamefont
  {Francica}}\ and\ \bibinfo {author} {\bibfnamefont {L.}~\bibnamefont
  {Dell{\textquotesingle}Anna}},\ }\bibfield  {title} {\bibinfo {title}
  {{Correlations, long-range entanglement, and dynamics in long-range Kitaev
  chains}},\ }\bibfield  {journal} {\bibinfo  {journal} {Physical Review B}\
  }\textbf {\bibinfo {volume} {106}},\ \href
  {https://doi.org/10.1103/physrevb.106.155126} {10.1103/physrevb.106.155126}
  (\bibinfo {year} {2022})\BibitemShut {NoStop}%
\bibitem [{\citenamefont {Gong}\ \emph {et~al.}(2016)\citenamefont {Gong},
  \citenamefont {Maghrebi}, \citenamefont {Hu}, \citenamefont {Wall},
  \citenamefont {Foss-Feig},\ and\ \citenamefont {Gorshkov}}]{Gong16}%
  \BibitemOpen
  \bibfield  {author} {\bibinfo {author} {\bibfnamefont {Z.-X.}\ \bibnamefont
  {Gong}}, \bibinfo {author} {\bibfnamefont {M.~F.}\ \bibnamefont {Maghrebi}},
  \bibinfo {author} {\bibfnamefont {A.}~\bibnamefont {Hu}}, \bibinfo {author}
  {\bibfnamefont {M.~L.}\ \bibnamefont {Wall}}, \bibinfo {author}
  {\bibfnamefont {M.}~\bibnamefont {Foss-Feig}},\ and\ \bibinfo {author}
  {\bibfnamefont {A.~V.}\ \bibnamefont {Gorshkov}},\ }\bibfield  {title}
  {\bibinfo {title} {Topological phases with long-range interactions},\
  }\bibfield  {journal} {\bibinfo  {journal} {Physical Review B}\ }\textbf
  {\bibinfo {volume} {93}},\ \href {https://doi.org/10.1103/physrevb.93.041102}
  {10.1103/physrevb.93.041102} (\bibinfo {year} {2016})\BibitemShut {NoStop}%
\bibitem [{\citenamefont {Lapa}\ and\ \citenamefont {Levin}(2021)}]{Lapa21}%
  \BibitemOpen
  \bibfield  {author} {\bibinfo {author} {\bibfnamefont {M.~F.}\ \bibnamefont
  {Lapa}}\ and\ \bibinfo {author} {\bibfnamefont {M.}~\bibnamefont {Levin}},\
  }\href {https://doi.org/10.48550/ARXIV.2107.11396} {\bibinfo {title}
  {Stability of ground state degeneracy to long-range interactions}} (\bibinfo
  {year} {2021})\BibitemShut {NoStop}%
\bibitem [{\citenamefont {Altland}\ and\ \citenamefont
  {Zirnbauer}(1997)}]{Altland97}%
  \BibitemOpen
  \bibfield  {author} {\bibinfo {author} {\bibfnamefont {A.}~\bibnamefont
  {Altland}}\ and\ \bibinfo {author} {\bibfnamefont {M.~R.}\ \bibnamefont
  {Zirnbauer}},\ }\bibfield  {title} {\bibinfo {title} {Nonstandard symmetry
  classes in mesoscopic normal-superconducting hybrid structures},\ }\href
  {https://doi.org/10.1103/PhysRevB.55.1142} {\bibfield  {journal} {\bibinfo
  {journal} {Phys. Rev. B}\ }\textbf {\bibinfo {volume} {55}},\ \bibinfo
  {pages} {1142} (\bibinfo {year} {1997})}\BibitemShut {NoStop}%
\bibitem [{\citenamefont {Gong}\ \emph {et~al.}(2023)\citenamefont {Gong},
  \citenamefont {Guaita},\ and\ \citenamefont {Cirac}}]{Gong22}%
  \BibitemOpen
  \bibfield  {author} {\bibinfo {author} {\bibfnamefont {Z.}~\bibnamefont
  {Gong}}, \bibinfo {author} {\bibfnamefont {T.}~\bibnamefont {Guaita}},\ and\
  \bibinfo {author} {\bibfnamefont {J.~I.}\ \bibnamefont {Cirac}},\ }\bibfield
  {title} {\bibinfo {title} {{Long-Range Free Fermions: Lieb-Robinson Bound,
  Clustering Properties, and Topological Phases}},\ }\href
  {https://doi.org/10.1103/PhysRevLett.130.070401} {\bibfield  {journal}
  {\bibinfo  {journal} {Phys. Rev. Lett.}\ }\textbf {\bibinfo {volume} {130}},\
  \bibinfo {pages} {070401} (\bibinfo {year} {2023})}\BibitemShut {NoStop}%
\bibitem [{Note1()}]{Note1}%
  \BibitemOpen
  \bibinfo {note} {While this is a physically different setting, the analysis
  is almost identical, see Appendix \ref {app:AIII}.}\BibitemShut {Stop}%
\bibitem [{\citenamefont {Su}\ \emph {et~al.}(1979)\citenamefont {Su},
  \citenamefont {Schrieffer},\ and\ \citenamefont {Heeger}}]{Su79}%
  \BibitemOpen
  \bibfield  {author} {\bibinfo {author} {\bibfnamefont {W.~P.}\ \bibnamefont
  {Su}}, \bibinfo {author} {\bibfnamefont {J.~R.}\ \bibnamefont {Schrieffer}},\
  and\ \bibinfo {author} {\bibfnamefont {A.~J.}\ \bibnamefont {Heeger}},\
  }\bibfield  {title} {\bibinfo {title} {Solitons in polyacetylene},\ }\href
  {https://doi.org/10.1103/PhysRevLett.42.1698} {\bibfield  {journal} {\bibinfo
   {journal} {Phys. Rev. Lett.}\ }\textbf {\bibinfo {volume} {42}},\ \bibinfo
  {pages} {1698} (\bibinfo {year} {1979})}\BibitemShut {NoStop}%
\bibitem [{\citenamefont {B{\"o}ttcher}\ and\ \citenamefont
  {Silbermann}(2006)}]{Boettcher06}%
  \BibitemOpen
  \bibfield  {author} {\bibinfo {author} {\bibfnamefont {A.}~\bibnamefont
  {B{\"o}ttcher}}\ and\ \bibinfo {author} {\bibfnamefont {B.}~\bibnamefont
  {Silbermann}},\ }\href@noop {} {\emph {\bibinfo {title} {Analysis of Toeplitz
  Operators}}},\ Springer Monographs in Mathematics\ (\bibinfo  {publisher}
  {Springer Berlin Heidelberg},\ \bibinfo {year} {2006})\BibitemShut {NoStop}%
\bibitem [{\citenamefont {McCoy}\ and\ \citenamefont {Wu}(1973)}]{McCoy73}%
  \BibitemOpen
  \bibfield  {author} {\bibinfo {author} {\bibfnamefont {B.~M.}\ \bibnamefont
  {McCoy}}\ and\ \bibinfo {author} {\bibfnamefont {T.~T.}\ \bibnamefont {Wu}},\
  }\href@noop {} {\emph {\bibinfo {title} {{The two-dimensional Ising
  model}}}}\ (\bibinfo  {publisher} {Harvard University Press},\ \bibinfo
  {year} {1973})\BibitemShut {NoStop}%
\bibitem [{Note2()}]{Note2}%
  \BibitemOpen
  \bibinfo {note} {The proof is given in Appendix \ref {app:Wiener-Hopf}. To
  exclude accidental edge modes we need results found in Refs.~\cite
  {Widom60,Douglas80,Boettcher06}.}\BibitemShut {Stop}%
\bibitem [{Note3()}]{Note3}%
  \BibitemOpen
  \bibinfo {note} {This shift is an application of an `SPT entangler' \cite
  {Jones21a,Tantivasadakarn23}.}\BibitemShut {Stop}%
\bibitem [{\citenamefont {Zygmund}(2002)}]{Zygmund02}%
  \BibitemOpen
  \bibfield  {author} {\bibinfo {author} {\bibfnamefont {A.}~\bibnamefont
  {Zygmund}},\ }\href@noop {} {\emph {\bibinfo {title} {Trigonometric
  series}}}\ (\bibinfo  {publisher} {Cambridge University Press},\ \bibinfo
  {year} {2002})\BibitemShut {NoStop}%
\bibitem [{\citenamefont {Grafakos}(2008)}]{Grafakos08}%
  \BibitemOpen
  \bibfield  {author} {\bibinfo {author} {\bibfnamefont {L.}~\bibnamefont
  {Grafakos}},\ }\href@noop {} {\emph {\bibinfo {title} {Classical Fourier
  Analysis}}}\ (\bibinfo  {publisher} {Springer},\ \bibinfo {year}
  {2008})\BibitemShut {NoStop}%
\bibitem [{\citenamefont {Balabanov}\ \emph {et~al.}(2021)\citenamefont
  {Balabanov}, \citenamefont {Erkensten},\ and\ \citenamefont
  {Johannesson}}]{Balabanov21}%
  \BibitemOpen
  \bibfield  {author} {\bibinfo {author} {\bibfnamefont {O.}~\bibnamefont
  {Balabanov}}, \bibinfo {author} {\bibfnamefont {D.}~\bibnamefont
  {Erkensten}},\ and\ \bibinfo {author} {\bibfnamefont {H.}~\bibnamefont
  {Johannesson}},\ }\bibfield  {title} {\bibinfo {title} {Topology of critical
  chiral phases: Multiband insulators and superconductors},\ }\bibfield
  {journal} {\bibinfo  {journal} {Physical Review Research}\ }\textbf {\bibinfo
  {volume} {3}},\ \href {https://doi.org/10.1103/physrevresearch.3.043048}
  {10.1103/physrevresearch.3.043048} (\bibinfo {year} {2021})\BibitemShut
  {NoStop}%
\bibitem [{{\relax DLMF}()}]{NIST:DLMF}%
  \BibitemOpen
  {\relax DLMF},\ \href {http://dlmf.nist.gov/} {\bibinfo {title} {{\it NIST
  Digital Library of Mathematical Functions}}},\ \bibinfo {howpublished}
  {http://dlmf.nist.gov/, Release 1.1.7 of 2022-10-15},\ \bibinfo {note}
  {{F.~W.~J. Olver, A.~B. {Olde Daalhuis}, D.~W. Lozier, B.~I. Schneider, R.~F.
  Boisvert, C.~W. Clark, B.~R. Miller, B.~V. Saunders, H.~S. Cohl, and M.~A.
  McClain, eds.}}\BibitemShut {Stop}%
\bibitem [{\citenamefont {Boettcher}\ and\ \citenamefont
  {Widom}(2006)}]{Boettcher06b}%
  \BibitemOpen
  \bibfield  {author} {\bibinfo {author} {\bibfnamefont {A.}~\bibnamefont
  {Boettcher}}\ and\ \bibinfo {author} {\bibfnamefont {H.}~\bibnamefont
  {Widom}},\ }\href {https://doi.org/10.48550/ARXIV.MATH/0604009} {\bibinfo
  {title} {{Szegö via Jacobi}}} (\bibinfo {year} {2006})\BibitemShut {NoStop}%
\bibitem [{Note4()}]{Note4}%
  \BibitemOpen
  \bibinfo {note} {There is a symmetric claim in the case with $\omega
  <0$.}\BibitemShut {Stop}%
\bibitem [{\citenamefont {Ekstr{\"o}m}\ \emph {et~al.}(2018)\citenamefont
  {Ekstr{\"o}m}, \citenamefont {Furci},\ and\ \citenamefont
  {Serra-Capizzano}}]{Ekstrom18}%
  \BibitemOpen
  \bibfield  {author} {\bibinfo {author} {\bibfnamefont {S.-E.}\ \bibnamefont
  {Ekstr{\"o}m}}, \bibinfo {author} {\bibfnamefont {I.}~\bibnamefont {Furci}},\
  and\ \bibinfo {author} {\bibfnamefont {S.}~\bibnamefont {Serra-Capizzano}},\
  }\bibfield  {title} {\bibinfo {title} {{Exact formulae and matrix-less
  eigensolvers for block banded symmetric Toeplitz matrices}},\ }\href@noop {}
  {\bibfield  {journal} {\bibinfo  {journal} {BIT Numerical Mathematics}\
  }\textbf {\bibinfo {volume} {58}},\ \bibinfo {pages} {937} (\bibinfo {year}
  {2018})}\BibitemShut {NoStop}%
\bibitem [{\citenamefont {Basor}\ \emph {et~al.}(2018)\citenamefont {Basor},
  \citenamefont {Dubail}, \citenamefont {Emig},\ and\ \citenamefont
  {Santachiara}}]{Basor18}%
  \BibitemOpen
  \bibfield  {author} {\bibinfo {author} {\bibfnamefont {E.}~\bibnamefont
  {Basor}}, \bibinfo {author} {\bibfnamefont {J.}~\bibnamefont {Dubail}},
  \bibinfo {author} {\bibfnamefont {T.}~\bibnamefont {Emig}},\ and\ \bibinfo
  {author} {\bibfnamefont {R.}~\bibnamefont {Santachiara}},\ }\bibfield
  {title} {\bibinfo {title} {{Modified Szegö{\textendash}Widom Asymptotics for
  Block Toeplitz Matrices with Zero Modes}},\ }\href
  {https://doi.org/10.1007/s10955-018-2177-8} {\bibfield  {journal} {\bibinfo
  {journal} {Journal of Statistical Physics}\ }\textbf {\bibinfo {volume}
  {174}},\ \bibinfo {pages} {28} (\bibinfo {year} {2018})}\BibitemShut
  {NoStop}%
\bibitem [{\citenamefont {Hartwig}\ and\ \citenamefont
  {Fisher}(1969)}]{Hartwig69}%
  \BibitemOpen
  \bibfield  {author} {\bibinfo {author} {\bibfnamefont {R.~E.}\ \bibnamefont
  {Hartwig}}\ and\ \bibinfo {author} {\bibfnamefont {M.~E.}\ \bibnamefont
  {Fisher}},\ }\bibfield  {title} {\bibinfo {title} {{Asymptotic behavior of
  Toeplitz matrices and determinants}},\ }\href
  {https://doi.org/10.1007/BF00247509} {\bibfield  {journal} {\bibinfo
  {journal} {Archive for Rational Mechanics and Analysis}\ }\textbf {\bibinfo
  {volume} {32}},\ \bibinfo {pages} {190} (\bibinfo {year} {1969})}\BibitemShut
  {NoStop}%
\bibitem [{\citenamefont {Widom}(1990)}]{Widom90}%
  \BibitemOpen
  \bibfield  {author} {\bibinfo {author} {\bibfnamefont {H.}~\bibnamefont
  {Widom}},\ }\bibfield  {title} {\bibinfo {title} {{Eigenvalue distribution of
  nonselfadjoint Toeplitz matrices and the asymptotics of Toeplitz determinants
  in the case of nonvanishing index}},\ }\href@noop {} {\bibfield  {journal}
  {\bibinfo  {journal} {Operator Theory: Advances and Applications}\ }\textbf
  {\bibinfo {volume} {48}},\ \bibinfo {pages} {387} (\bibinfo {year}
  {1990})}\BibitemShut {NoStop}%
\bibitem [{\citenamefont {Jones}\ and\ \citenamefont
  {Verresen}(2019)}]{Jones19}%
  \BibitemOpen
  \bibfield  {author} {\bibinfo {author} {\bibfnamefont {N.~G.}\ \bibnamefont
  {Jones}}\ and\ \bibinfo {author} {\bibfnamefont {R.}~\bibnamefont
  {Verresen}},\ }\bibfield  {title} {\bibinfo {title} {{Asymptotic Correlations
  in Gapped and Critical Topological Phases of 1D Quantum Systems}},\ }\href
  {https://doi.org/10.1007/s10955-019-02257-9} {\bibfield  {journal} {\bibinfo
  {journal} {Journal of Statistical Physics}\ }\textbf {\bibinfo {volume}
  {175}},\ \bibinfo {pages} {1164} (\bibinfo {year} {2019})}\BibitemShut
  {NoStop}%
\bibitem [{\citenamefont {Verresen}(2020)}]{Verresen20}%
  \BibitemOpen
  \bibfield  {author} {\bibinfo {author} {\bibfnamefont {R.}~\bibnamefont
  {Verresen}},\ }\href {https://doi.org/10.48550/ARXIV.2003.05453} {\bibinfo
  {title} {Topology and edge states survive quantum criticality between
  topological insulators}} (\bibinfo {year} {2020})\BibitemShut {NoStop}%
\bibitem [{\citenamefont {Widom}(1960)}]{Widom60}%
  \BibitemOpen
  \bibfield  {author} {\bibinfo {author} {\bibfnamefont {H.}~\bibnamefont
  {Widom}},\ }\bibfield  {title} {\bibinfo {title} {{Inversion of Toeplitz
  matrices II}},\ }\href {https://doi.org/10.1215/ijm/1255455736} {\bibfield
  {journal} {\bibinfo  {journal} {Illinois Journal of Mathematics}\ }\textbf
  {\bibinfo {volume} {4}},\ \bibinfo {pages} {88 } (\bibinfo {year}
  {1960})}\BibitemShut {NoStop}%
\bibitem [{\citenamefont {Douglas}(1980)}]{Douglas80}%
  \BibitemOpen
  \bibfield  {author} {\bibinfo {author} {\bibfnamefont {R.~G.}\ \bibnamefont
  {Douglas}},\ }\href@noop {} {\emph {\bibinfo {title} {Banach algebra
  techniques in the theory of Toeplitz operators}}},\ Vol.~\bibinfo {volume}
  {15}\ (\bibinfo  {publisher} {American Mathematical Soc.},\ \bibinfo {year}
  {1980})\BibitemShut {NoStop}%
\bibitem [{\citenamefont {Jones}\ \emph {et~al.}(2021)\citenamefont {Jones},
  \citenamefont {Bibo}, \citenamefont {Jobst}, \citenamefont {Pollmann},
  \citenamefont {Smith},\ and\ \citenamefont {Verresen}}]{Jones21a}%
  \BibitemOpen
  \bibfield  {author} {\bibinfo {author} {\bibfnamefont {N.~G.}\ \bibnamefont
  {Jones}}, \bibinfo {author} {\bibfnamefont {J.}~\bibnamefont {Bibo}},
  \bibinfo {author} {\bibfnamefont {B.}~\bibnamefont {Jobst}}, \bibinfo
  {author} {\bibfnamefont {F.}~\bibnamefont {Pollmann}}, \bibinfo {author}
  {\bibfnamefont {A.}~\bibnamefont {Smith}},\ and\ \bibinfo {author}
  {\bibfnamefont {R.}~\bibnamefont {Verresen}},\ }\bibfield  {title} {\bibinfo
  {title} {Skeleton of matrix-product-state-solvable models connecting
  topological phases of matter},\ }\bibfield  {journal} {\bibinfo  {journal}
  {Physical Review Research}\ }\textbf {\bibinfo {volume} {3}},\ \href
  {https://doi.org/10.1103/physrevresearch.3.033265}
  {10.1103/physrevresearch.3.033265} (\bibinfo {year} {2021})\BibitemShut
  {NoStop}%
\bibitem [{\citenamefont {Tantivasadakarn}\ \emph {et~al.}(2023)\citenamefont
  {Tantivasadakarn}, \citenamefont {Thorngren}, \citenamefont {Vishwanath},\
  and\ \citenamefont {Verresen}}]{Tantivasadakarn23}%
  \BibitemOpen
  \bibfield  {author} {\bibinfo {author} {\bibfnamefont {N.}~\bibnamefont
  {Tantivasadakarn}}, \bibinfo {author} {\bibfnamefont {R.}~\bibnamefont
  {Thorngren}}, \bibinfo {author} {\bibfnamefont {A.}~\bibnamefont
  {Vishwanath}},\ and\ \bibinfo {author} {\bibfnamefont {R.}~\bibnamefont
  {Verresen}},\ }\bibfield  {title} {\bibinfo {title} {{Pivot Hamiltonians as
  generators of symmetry and entanglement}},\ }\bibfield  {journal} {\bibinfo
  {journal} {{SciPost} Physics}\ }\textbf {\bibinfo {volume} {14}},\ \href
  {https://doi.org/10.21468/scipostphys.14.2.012}
  {10.21468/scipostphys.14.2.012} (\bibinfo {year} {2023})\BibitemShut
  {NoStop}%
\bibitem [{\citenamefont {Olver}(1974)}]{Olver97}%
  \BibitemOpen
  \bibfield  {author} {\bibinfo {author} {\bibfnamefont {F.~W.~J.}\
  \bibnamefont {Olver}},\ }\href@noop {} {\emph {\bibinfo {title} {Asymptotics
  and Special Functions}}}\ (\bibinfo  {publisher} {New York; London: Academic
  Press},\ \bibinfo {year} {1974})\BibitemShut {NoStop}%
\bibitem [{\citenamefont {Temme}(2014)}]{Temme14}%
  \BibitemOpen
  \bibfield  {author} {\bibinfo {author} {\bibfnamefont {N.~M.}\ \bibnamefont
  {Temme}},\ }\href@noop {} {\emph {\bibinfo {title} {Asymptotic methods for
  integrals}}},\ Vol.~\bibinfo {volume} {6}\ (\bibinfo  {publisher} {World
  Scientific},\ \bibinfo {year} {2014})\BibitemShut {NoStop}%
\bibitem [{\citenamefont {Miller}(2006)}]{Miller2006}%
  \BibitemOpen
  \bibfield  {author} {\bibinfo {author} {\bibfnamefont {P.~D.}\ \bibnamefont
  {Miller}},\ }\href@noop {} {\emph {\bibinfo {title} {Applied Asymptotic
  Analysis}}},\ \bibinfo {series} {Graduate Studies in Mathematics},
  Vol.~\bibinfo {volume} {75}\ (\bibinfo  {publisher} {American Mathematical
  Society},\ \bibinfo {year} {2006})\BibitemShut {NoStop}%
\bibitem [{\citenamefont {Fendley}(2014)}]{Fendley14}%
  \BibitemOpen
  \bibfield  {author} {\bibinfo {author} {\bibfnamefont {P.}~\bibnamefont
  {Fendley}},\ }\bibfield  {title} {\bibinfo {title} {Free parafermions},\
  }\href {https://doi.org/10.1088/1751-8113/47/7/075001} {\bibfield  {journal}
  {\bibinfo  {journal} {Journal of Physics A: Mathematical and Theoretical}\
  }\textbf {\bibinfo {volume} {47}},\ \bibinfo {pages} {075001} (\bibinfo
  {year} {2014})}\BibitemShut {NoStop}%
\bibitem [{\citenamefont {Dai}\ \emph {et~al.}(2009)\citenamefont {Dai},
  \citenamefont {Geary},\ and\ \citenamefont {Kadanoff}}]{Dai09}%
  \BibitemOpen
  \bibfield  {author} {\bibinfo {author} {\bibfnamefont {H.}~\bibnamefont
  {Dai}}, \bibinfo {author} {\bibfnamefont {Z.}~\bibnamefont {Geary}},\ and\
  \bibinfo {author} {\bibfnamefont {L.~P.}\ \bibnamefont {Kadanoff}},\
  }\bibfield  {title} {\bibinfo {title} {{Asymptotics of eigenvalues and
  eigenvectors of Toeplitz matrices}},\ }\href
  {https://doi.org/10.1088/1742-5468/2009/05/p05012} {\bibfield  {journal}
  {\bibinfo  {journal} {Journal of Statistical Mechanics: Theory and
  Experiment}\ }\textbf {\bibinfo {volume} {2009}},\ \bibinfo {pages} {P05012}
  (\bibinfo {year} {2009})}\BibitemShut {NoStop}%
\bibitem [{\citenamefont {Kadanoff}(2009)}]{Kadanoff09}%
  \BibitemOpen
  \bibfield  {author} {\bibinfo {author} {\bibfnamefont {L.~P.}\ \bibnamefont
  {Kadanoff}},\ }\href {https://doi.org/10.48550/ARXIV.0906.0760} {\bibinfo
  {title} {{Expansions for Eigenfunction and Eigenvalues of large-$n$ Toeplitz
  Matrices}}} (\bibinfo {year} {2009})\BibitemShut {NoStop}%
\bibitem [{\citenamefont {Böttcher}\ \emph {et~al.}(2017)\citenamefont
  {Böttcher}, \citenamefont {Bogoya}, \citenamefont {Grudsky},\ and\
  \citenamefont {Maximenko}}]{Boettcher17}%
  \BibitemOpen
  \bibfield  {author} {\bibinfo {author} {\bibfnamefont {A.}~\bibnamefont
  {Böttcher}}, \bibinfo {author} {\bibfnamefont {J.~M.}\ \bibnamefont
  {Bogoya}}, \bibinfo {author} {\bibfnamefont {S.~M.}\ \bibnamefont
  {Grudsky}},\ and\ \bibinfo {author} {\bibfnamefont {E.~A.}\ \bibnamefont
  {Maximenko}},\ }\bibfield  {title} {\bibinfo {title} {Asymptotics of
  eigenvalues and eigenvectors of toeplitz matrices},\ }\href
  {https://doi.org/10.1070/SM8865} {\bibfield  {journal} {\bibinfo  {journal}
  {Sbornik: Mathematics}\ }\textbf {\bibinfo {volume} {208}},\ \bibinfo {pages}
  {1578} (\bibinfo {year} {2017})}\BibitemShut {NoStop}%
\bibitem [{Note5()}]{Note5}%
  \BibitemOpen
  \bibinfo {note} {Despite the leading notation, $b_-(1/z)$ is continuous and
  not necessarily analytic on the unit circle; hence this statement is
  non-trivial. It follows from Pollard's theorem (a generalisation of Cauchy's
  theorem for functions analytic inside and continuous on a simple closed
  contour) \cite {McCoy73}.}\BibitemShut {Stop}%
\bibitem [{\citenamefont {Andersson}\ \emph {et~al.}(1999)\citenamefont
  {Andersson}, \citenamefont {Boman},\ and\ \citenamefont
  {\"Ostlund}}]{Andersson99}%
  \BibitemOpen
  \bibfield  {author} {\bibinfo {author} {\bibfnamefont {M.}~\bibnamefont
  {Andersson}}, \bibinfo {author} {\bibfnamefont {M.}~\bibnamefont {Boman}},\
  and\ \bibinfo {author} {\bibfnamefont {S.}~\bibnamefont {\"Ostlund}},\
  }\bibfield  {title} {\bibinfo {title} {Density-matrix renormalization group
  for a gapless system of free fermions},\ }\href
  {https://doi.org/10.1103/PhysRevB.59.10493} {\bibfield  {journal} {\bibinfo
  {journal} {Phys. Rev. B}\ }\textbf {\bibinfo {volume} {59}},\ \bibinfo
  {pages} {10493} (\bibinfo {year} {1999})}\BibitemShut {NoStop}%
\bibitem [{\citenamefont {Wood}(1992)}]{Wood92}%
  \BibitemOpen
  \bibfield  {author} {\bibinfo {author} {\bibfnamefont {D.}~\bibnamefont
  {Wood}},\ }\href {http://www.cs.kent.ac.uk/pubs/1992/110} {\emph {\bibinfo
  {title} {The Computation of Polylogarithms}}},\ \bibinfo {type} {Tech. Rep.}\
  \bibinfo {number} {15-92*}\ (\bibinfo  {institution} {University of Kent,
  Computing Laboratory},\ \bibinfo {address} {University of Kent, Canterbury,
  UK},\ \bibinfo {year} {1992})\BibitemShut {NoStop}%
\bibitem [{Note6()}]{Note6}%
  \BibitemOpen
  \bibinfo {note} {More carefully: Proposition \ref {prop:asymptotics} and
  Theorem \ref {thm:HartwigFisher} give singularity-filling for Toeplitz
  determinants up to $\omega =3$. If we additionally assume that the spectrum
  is such that the edge mode energies correspond to the Toeplitz determinant
  according to the inductive method outlined in the main text, then we have
  singularity filling for the splittings (Conjecture \ref
  {conjecture}).}\BibitemShut {Stop}%
\bibitem [{\citenamefont {Devinatz}(1964)}]{Devinatz}%
  \BibitemOpen
  \bibfield  {author} {\bibinfo {author} {\bibfnamefont {A.}~\bibnamefont
  {Devinatz}},\ }\bibfield  {title} {\bibinfo {title} {{Toeplitz Operators on
  $H^2$ Spaces}},\ }\href {http://www.jstor.org/stable/1994297} {\bibfield
  {journal} {\bibinfo  {journal} {Transactions of the American Mathematical
  Society}\ }\textbf {\bibinfo {volume} {112}},\ \bibinfo {pages} {304}
  (\bibinfo {year} {1964})}\BibitemShut {NoStop}%
\bibitem [{Note7()}]{Note7}%
  \BibitemOpen
  \bibinfo {note} {There is a misprint in the statement of this lemma in \cite
  {Widom90}.}\BibitemShut {Stop}%
\bibitem [{\citenamefont {Verresen}\ \emph {et~al.}(2021)\citenamefont
  {Verresen}, \citenamefont {Thorngren}, \citenamefont {Jones},\ and\
  \citenamefont {Pollmann}}]{Verresen21}%
  \BibitemOpen
  \bibfield  {author} {\bibinfo {author} {\bibfnamefont {R.}~\bibnamefont
  {Verresen}}, \bibinfo {author} {\bibfnamefont {R.}~\bibnamefont {Thorngren}},
  \bibinfo {author} {\bibfnamefont {N.~G.}\ \bibnamefont {Jones}},\ and\
  \bibinfo {author} {\bibfnamefont {F.}~\bibnamefont {Pollmann}},\ }\bibfield
  {title} {\bibinfo {title} {{Gapless Topological Phases and Symmetry-Enriched
  Quantum Criticality}},\ }\bibfield  {journal} {\bibinfo  {journal} {Physical
  Review X}\ }\textbf {\bibinfo {volume} {11}},\ \href
  {https://doi.org/10.1103/physrevx.11.041059} {10.1103/physrevx.11.041059}
  (\bibinfo {year} {2021})\BibitemShut {NoStop}%
\bibitem [{\citenamefont {Lang}(2003)}]{Lang03}%
  \BibitemOpen
  \bibfield  {author} {\bibinfo {author} {\bibfnamefont {S.}~\bibnamefont
  {Lang}},\ }\href@noop {} {\emph {\bibinfo {title} {Complex analysis}}},\
  Vol.\ \bibinfo {volume} {103}\ (\bibinfo  {publisher} {Springer Science \&
  Business Media},\ \bibinfo {year} {2003})\BibitemShut {NoStop}%
\bibitem [{\citenamefont {Jones}\ and\ \citenamefont
  {Verresen}(2021)}]{Jones21b}%
  \BibitemOpen
  \bibfield  {author} {\bibinfo {author} {\bibfnamefont {N.~G.}\ \bibnamefont
  {Jones}}\ and\ \bibinfo {author} {\bibfnamefont {R.}~\bibnamefont
  {Verresen}},\ }\href {https://doi.org/10.48550/ARXIV.2105.13359} {\bibinfo
  {title} {Exact correlations in topological quantum chains}} (\bibinfo {year}
  {2021})\BibitemShut {NoStop}%
\end{thebibliography}%

\appendix
\onecolumngrid
\section{The BDI chain}
\label{app:solution}
Let us consider the BDI class of translation invariant spinless free-fermions with time-reversal symmetry on a chain of length $L$:
\begin{align}
H_{\mathrm{BDI}} = \frac{\rmi}{2}  \sum_{m,n =0}^{L-1} \tau_{m-n} \tilde \gamma_n \gamma_{m}.  \end{align}
Take an infinite set of real coupling coefficients $\{t_n\}_{n\in\mathbb{Z}}$; then we can straightforwardly define an open chain by putting $\tau_{m-n} = t_{m-n}$. 

We can define the corresponding closed chain by writing 
\begin{align}
\tau_{m-n} = \begin{cases}
t_{m-n} \qquad   -L/2\leq m-n < L/2 \\
t_{m-n+L} \qquad   m-n < -L/2 \\
t_{m-n-L} \qquad   L/2 \leq m-n,
\end{cases}
\end{align}
this choice is considered in Ref.~\cite{Basor18}. (A less general case, where $t_{m-n}$ depends on $\lvert m-n\rvert$ is considered in many works on the long-range Kitaev chain, e.g., see discussion in Refs.~\cite{Vodola14,Alecce17}). To solve this rigorously for finite $L$ we should impose either periodic or anti-periodic boundary conditions for the fermions and proceed. However, intuitively, in the thermodynamic limit $L\rightarrow\infty$, the effects of couplings that wrap around the chain will vanish algebraically with system size. We will follow Ref.~\cite{Defenu21} and instead consider a sequence of finite-range chains, where we truncate:
\begin{align}
\tau_{m-n} = \begin{cases}
t_{m-n} \qquad&   -L/2\leq m-n < L/2 \\
0\qquad   &\textrm{otherwise.}
\end{cases}
\end{align}
Then we can use the usual method of solving such chains via Fourier transformation and Bogoliubov transformation (see e.g., Ref.~\cite{Verresen18}), leading to the complex function $f(z)$. As a consequence of the absolute-summability, in the thermodynamic limit we can write continuous functions $\eps_k$ and $\varphi_k$ such that on the unit circle $f(\rme^{\rmi k})=\eps_k \rme^{\rmi \varphi_k}$. 
Then the Hamiltonian has the diagonal form:
$H_{\mathrm{BDI}} = -\sum_k \eps_k d^{\dagger}_kd^{\vphantom{\dagger}}_k$ where $\eps_k = \lvert f(\rme^{\rmi k})\rvert$ and 
\begin{align}
    \left(\begin{array}{c}d_k \\d^\dagger_k\end{array}\right)=\rme^{\rmi \varphi_k \sigma_x/2} \left(\begin{array}{c}c_k \\c^\dagger_{-k}\end{array}\right).
\end{align} 
In this expression, $c_k=\sum_n \rme^{-\rmi k n}c_n$ are the Fourier transformed spinless fermion operators $c_n=(\gamma_n+\rmi \tilde\gamma_n)/2$.
\section{Singularities of $f(z)$ and related functions}
\subsection{Defining singularities via Fourier coefficients}
In the main text we introduce the idea of singularity-filling, for singularities of certain functions defined on the unit circle in the complex plane. These singularities correspond to particular momenta $0\leq k_s < 2\pi$, or, equivalently, to points on the unit circle $\rme^{\rmi k_s}$. In general, we say that
a function $h(z)$ on the unit circle has singularities at $\{k_s\}_{1\leq s\leq r}$ if it has an asymptotic Fourier expansion of the form 
$h_n =\sum_{s=1}^r \rme^{\rmi n k_s }n^{-\Omega_{k_s}} \left( a_s + o(1) \right)$. In the main text we suppose that the $o(1)$ terms are all of the form $n^{-b}$ for $b>0$. One may consider generalisations, such as allowing terms of the form $\log(n)^a n^{-b}$, for some $a \in \mathbb{Z}$ and $b>0$. We explain below that the proof of Theorem \ref{thm2} can accommodate this particular generalisation.

This definition of singularity, based on Fourier coefficients, can be related to other notions of analytical singularity. For example, suppose $h(z)$ has branch point(s) on the unit circle at $\rme^{\rmi k_s}$, and that we can analytically continue the function to the complex plane up to some branch cuts. When we compute asymptotic Fourier coefficients, we are dominated by integrals near the branch points. Then, supposing an appropriate expansion at the singularity, using Watson's lemma \cite{Olver97,Temme14} we find a dominant contribution (at each singularity) of the form $\rme^{\rmi k_s n} n^{-\Omega_{k_s}}$ (a particular example of this is the calculation following Eq.~\ref{eq:fouriercontour} below). Another relevant definition of singularity is a discontinuity in some derivative of the function $h(z)$. More precisely, we can characterise the smoothness of the function $h(z)$ according to the number of continuous derivatives. Then we have well-known results relating this smoothness to the asymptotic decay of the Fourier coefficients; see, for example, Ref.~\cite{Grafakos08}.  

Since our results in the main text depend directly on certain Fourier coefficients, we choose to use this definition of singularity for clarity. In analysing a particular problem with a chain corresponding to a function $f(z)$ that has some analytical singularity, one needs to then justify how this is reflected in the asymptotic Fourier expansion. Whether this is straightforward depends upon the particular choice of $f(z)$, but there are many general results available \cite{Grafakos08,Boettcher06}.
\subsection{Relationship between singularities of $f(z)$ and the Wiener-Hopf factors}
Our main results depend on several different, but closely related, functions. The function $f(z) = \sum_{n=-\infty}^\infty t_n z^n$ corresponds directly to the Hamiltonian, and the dominant asymptotic decay of the Fourier coefficients of $f(z)$ tells us the algebraic decay of the coupling coefficients. Note that $f(z)$ and $z^k f(z)$ necessarily have the same singularities, since we simply shift $t_n \rightarrow t_{n-k}$ and this does not change the asymptotic Fourier coefficients.

For $\omega>0$ [$\omega<0$] the edge mode wavefunctions depend on Fourier coefficients of the inverse Wiener-Hopf factor $b_-(1/z)^{-1}$ [$b_+(z)^{-1}$]; where $f(z) = z^\omega f_0(z)$ for $f_0(z)= b_+(z) b_-(z)$. Moreover, based on the Toeplitz determinant theory that underlies Conjecture \ref{conjecture}, the edge-mode splittings depend on the asymptotic Fourier coefficients of $m(z)=b_+(1/z)/b_-(1/z)$ [$l(z) =b_-(z)/b_+(z)]$ (this is explained in greater detail in Appendix \ref{app:filling}). These functions are clearly closely related, and this can be made quantitative. 

Let us then consider an $(\alpha>1)$-decaying Hamiltonian, with corresponding $f(z)$. These functions $f(z)$ are a subset of the class considered in Ref.~\cite{Hartwig69}, and we can make a corresponding analysis. Following Ref.~\cite{Hartwig69}, denote the Fourier coefficients of $b_+(z)$ by $r_n$, and the Fourier coefficients of $b_-(1/z)^{-1}$ by $q_n$. Then, from the Wiener-Hopf decomposition, we have that $r_0=q_0 =1$ and $r_{-m} = q_{-m} =0$ for any $m\in\mathbb{N}$. Moreover, $r_n$ and $q_n$ are absolutely summable and we have:
\begin{align}
m_n = \sum_{j=n}^\infty q_j r_{j-n} . \label{eq:convolution}
\end{align} 
Similarly denote the Fourier coefficients of $f_0(1/z)^{-1}$ by $s_n$ (which is in general doubly-infinite). Then we have:
\begin{align}
q_n = \sum_{j=n}^\infty s_j r_{j-n} . \label{eq:convolution2}
\end{align} 
Note that this calculation is exact for $1/f_0(1/z)$, but if we instead take $1/f(1/z)= z^{-\omega}/f_0(1/z)$ the Fourier coefficients are simply shifted. 

Finally, $b_+(z)^2$ has the same properties as $b_+(z)$, with Fourier coefficients $\tilde{r}_n$. We can then write
\begin{align}
m_n = \sum_{j=n}^\infty s_j \tilde{r}_{j-n} . \label{eq:convolution3}
\end{align} 

Let us analyse \eqref{eq:convolution2}, with analogous conclusions holding in the other cases. Suppose, as in the main text, that we have an asymptotic expansion for $s_n$ with certain singularities
\begin{align}
s_n &= \sum_{s=1}^r  a_s \rme^{\rmi n k_s }n^{-\Omega_{k_s}} (1+o(1)) ,
\end{align}
where we say the $o(1)$ term is `nice', i.e., that each term is an inverse power of $n$ (we may also have further subdominant terms that decay faster than any power of $n$, we will usually suppress them below). 
Now, using the absolute summability of the $r_n$, we see that $q_n$ has an asymptotic expansion with identical singularities $\{k_s\}$ (and corresponding orders $\{\Omega_{k_s}\}$) to $s_n$, we simply renormalise the coefficients in the expansion.
\begin{align}
q_n &= \sum_{s=1}^r \rme^{\rmi n k_s }n^{-\Omega_{k_s}} \left( a_s\left(\sum_{j=0}^{\infty} \rme^{\rmi k_s j} r_j \left( \frac{n}{n+j}\right)^{\Omega_{k_s}}  \right) + o(1) \right)\nonumber \\
&=  \sum_{s=1}^r \rme^{\rmi n k_s }n^{-\Omega_{k_s}} \left( a_s\left(b_+(\rme^{\rmi k_s})+o(1) \right) + o(1) \right)\label{eq:qn}
\end{align}
To justify that the orders of the singularities are the same, note that
since $f(z)$ corresponds to a gapped Hamiltonian, $b_+(z)$ cannot vanish on the unit circle.

For our purposes in Theorem \ref{thm2} and Conjecture \ref{conjecture}, we also want the $o(1)$ terms here to have a nice dependence on $n$. We now show that the expansion is in inverse powers of $n$ up to some cut off that
depends on $\Omega_{\textrm{min}}$, the dominant singularity. In particular, we will now show that 
\begin{align}
q_n 
&=  \sum_{s=1}^r \rme^{\rmi n k_s }n^{-\Omega_{k_s}} \left( a_s\left(b_+(\rme^{\rmi k}) + \sum_{j=1}^{\delta_0-1}A_j n^{-j} \right) + O(n^{-\delta_0}) \right)
\label{eq:singularities}\end{align}
for some known constants $A_j$, and $\delta_0 = \lfloor \Omega_{\textrm{min}} -1 \rfloor$.
The same conclusion will hold for $m_n$, by analogous calculations. 

To prove this, we first recall that functions in the class $C^\beta$ have $n=\lfloor \beta \rfloor$ continuous derivatives. 
Moreover, our assumption on the singularities of $1/f(1/z)$, where $\Omega_{\textrm{min}}$ is the order of the dominant singularity, implies the following. $\{f(z)^{\pm 1}, b_+(z)^{\pm 1}, b_-(z)^{\pm1}  \}$ are all in $ C^{\delta}$ on the unit circle for $ \lfloor \Omega_{\textrm{min}}-1 \rfloor <\delta<\Omega_{\textrm{min}}-1$ \cite{Boettcher06b}, and in particular have $\delta_0$ continuous derivatives.
Let us now revisit the crucial term in the expansion of $q_n$:
\begin{align}
\sum_{j=0}^{\infty} \rme^{\rmi k_s j} r_j \left( \frac{n}{n+j}\right)^{\Omega_{k_s}} &= \left( \sum_{j=0}^{\infty} \rme^{\rmi k_s j} r_j n^{\Omega_{k_s}}\frac{1}{\Gamma(\Omega_{k_s})}\int_0^\infty \rme^{-nt} \rme^{- jt }t^{\Omega_{k_s}-1}\rmd t\right)\nonumber \\&=n^{\Omega_{k_s}}\frac{1}{\Gamma(\Omega_{k_s})}\int_0^\infty b_+(\rme^{\rmi k_s} \rme^{-t})\rme^{-nt} t^{\Omega_{k_s}-1}\rmd t.
\end{align}
This is in a form amenable to Watson's lemma \cite{Olver97,Miller2006}, leading to the following asymptotic expansion:
\begin{align}
\sum_{j=0}^{\infty} \rme^{\rmi k_s j} r_j \left( \frac{n}{n+j}\right)^{\Omega_{k_s}} &=\sum_{k=0}^{\delta_0-1} A_k n^{-k} + O(n^{-\delta_0}),
\end{align}
for some constant coefficients $A_k$ that depend on derivatives of $b_+(z)$. This establishes \eqref{eq:singularities} above.

\section{Analysis of edge modes}

\subsection{Proof of Theorem \ref{thm1}}\label{app:Wiener-Hopf}
Consider the BDI Hamiltonian with open boundaries. In general \cite{Fendley14}, we can diagonalise by finding raising and lowering operators $\chi_\eps = \sum_{n=0}^{L-1} a_n \gamma_n+b_n \tilde\gamma_n$ that satisfy $[H_{\textrm{BDI}},\chi_\eps] = 2\eps \chi_\eps$. Evaluating the commutator reduces to the mathematical problem of finding the eigenvectors of a  \emph{block Toeplitz matrix}, for which analytical results are available only in special cases. Note that \emph{Toeplitz matrices} are matrices that are constant along diagonals. These constants are determined as Fourier coefficients $t_n$ of a generating function $t(z)$. A {block Toeplitz matrix} has the same structure as a (scalar) Toeplitz matrix, but the scalar constants on each diagonal are replaced by constant matrices of fixed size. 

One such solvable  case is that of exact zero modes; then $\eps =0$, and the problem reduces to finding eigenvectors in the kernel of scalar Toeplitz matrices. There do exist a variety of results for asymptotic behaviour of eigenvectors of scalar Toeplitz matrices \cite{Dai09,Kadanoff09}; for a review of the field see \cite{Boettcher17}. However, for topological Majorana zero modes, the splitting is generically exactly zero only in the infinite system size limit.
\subsubsection{Wiener-Hopf sum equations}

In their textbook on the Ising model \cite{McCoy73}, McCoy and Wu solve the following Wiener-Hopf sum equation:
\begin{align}
\sum_{m=0}^\infty c_{n-m} x_m = y_n \qquad n\geq 0,\label{sumeq}
\end{align}
subject to the condition $\sum_{n \in \mathbb{Z}} |c_n | < \infty$, and solutions are sought with bounded norm, i.e., $\sum_{n \in \mathbb{Z}} |x_n | < \infty.$ 

For our application, $y_n =0$, and we give the results for that case. Define $c(z) = \sum_{n \in \mathbb{Z}} c_n z^n$. Then, assuming $c(z)$ does not vanish on unit circle, we have the Wiener-Hopf decomposition $c(z) = z^\nu \beta_+(z) \beta_0 \beta_-(z)$, and let us  fix the overall normalisation so that $\beta_0=1$. Note this decomposition exists and each function has an absolutely convergent Fourier series due to the Weiner-L\'evy theorem \cite{McCoy73,Zygmund02}.

The general solutions of \eqref{sumeq} for $y=0$ are as follows:
\begin{align}
x_n = \begin{cases} 0 \qquad  &\nu \geq 0 \\
\sum_{m=0}^{|\nu| -1}a_m \left( \frac{1}{\beta_+(z)}\right)_{n-m} \qquad &\nu<0
\end{cases}.
\end{align}
 We thus see that for $\nu \geq 0$ there are no non-trivial solutions, while for $\nu <0$ we have $|\nu|$ solutions. Note the fixed chirality of this problem, it is always negative winding allowing solutions.
\subsubsection{Application to edge modes}
Consider the half-infinite OBC Hamiltonian  $H_{\mathrm{BDI}} =\rmi \sum_{n \geq 0,m\geq 0} t_{m-n}  \tilde\gamma_n \gamma_{m}$, assuming that $\sum_{n=-\infty}^\infty \lvert t_n\rvert <\infty$. For real chiral edge modes we have $\gamma_L = \sum_{n\geq 0} \alpha_n \gamma_n$. These satisfy $[H_{\mathrm{BDI}},\gamma_L]=0$. Calculating this commutator we find:
\begin{align}
[H_{\mathrm{BDI}},\gamma_L]&= 2\rmi \sum_{n,m,r \geq 0} t_{m-n}  \tilde\gamma_n \delta_{m,r}\alpha_r \nonumber\\&=2\rmi \sum_{n \geq 0}\sum_{m\geq0}\left( t_{m-n}   \alpha_m \right)\tilde\gamma_n\nonumber \\&=2\rmi \sum_{n \geq 0}\sum_{m\geq0}\left( \tilde{t}_{n-m}   \alpha_m \right)\tilde\gamma_n,
\end{align}
where $\tilde{t}_\alpha = t_{-\alpha}$. We thus see that if this commutator vanishes, then the $\alpha_m$ must be solutions to \eqref{sumeq}, for the choice $c(z) = \sum_{\alpha} \tilde{t}_\alpha z^\alpha = f(1/z)$. Suppose that $f(z) = z^\omega b_+(z) b_-(z)$, then we have that $c(z) = z^{-\omega} b_+(1/z) b_-(1/z)$. By considering dependence on $z$ and $1/z$ we have $b_\mp(1/z)=\beta_\pm(z)$, we reach the first part of Theorem \ref{thm1}. Note that to prove the independence of the edge modes, we use $b_-(1/z)^{-1} =\rme^{\sum_{k=1}^\infty V_{- k} z^{ k}}$ has all negative Fourier coefficients equal to zero \footnote{Despite the leading notation, $b_-(1/z)$ is continuous and not necessarily analytic on the unit circle; hence this statement is non-trivial. It follows from Pollard's theorem (a generalisation of Cauchy's theorem for functions analytic inside and continuous on a simple closed contour) \cite{McCoy73}.}.
The second part of Theorem \ref{thm1} follows straightforwardly by considering the inverted chain. Then we take $f(z)\rightarrow f(1/z)$ and switch $\gamma$ and $\tilde\gamma$. Alternatively one can repeat the above Wiener-Hopf calculation after inserting the ansatz for an imaginary chiral edge mode. 

As an aside, suppose we did not restrict to chiral zero modes, and take the following ansatz  $\chi_L = \sum_n (\alpha_n \gamma_n +\beta_n \tilde\gamma_n)$. Then calculating the commutator gives two independent problems of the form \eqref{sumeq}, one for $f(z)$ and one for $f(1/z)$. Hence, depending on winding number, at least one of them will have no non-trivial solutions and we are back in the chiral case.

Note that a Majorana edge mode is normalisable if $\sum_{n\geq0} \lvert g_n\rvert^2 <\infty$. The proof of Theorem \ref{thm1} leads to a stronger conclusion than stated: in fact the edge modes given are \emph{the only edge modes that exist} satisfying the condition $\sum_{n\geq0} \lvert g_n\rvert <\infty$. Hence, the discussion above based on results of \cite{McCoy73} does not immediately exclude `accidental' (non-topological) localised edged modes that are sufficiently delocalised that $\sum_{n\geq0} \lvert g_n\rvert \rightarrow \infty$. 
However, appealing to general results \cite{Widom60,Douglas80,Boettcher06} on invertibility of Toeplitz operators (over the sequence space $l^2$) leads to the conclusion that Theorem \ref{thm1} does indeed give us all of the Majorana edge modes.
\subsection{Proof of Theorem \ref{thm2}}\label{app:thm2} 
Here we prove a stronger form of Theorem \ref{thm2} given in the main text. The version in the main text is simpler to state, and follows from:
\begin{theorem}\label{thm:2}
 Consider a model corresponding to $f(z) = z^\omega b_+(z)b_-(z)$ with open boundary conditions, and suppose that the Fourier coefficients of $b_-(1/z)^{-1}$ have an expansion \begin{align}g_n =  \sum_{s=1}^r \left(\sum_{p=0}^{p_s^\star}\sum_{\Delta p \in P}\sum_{q =0}^{q_s^\star}  \rme^{\rmi n k_s } a_{s,p,q} \log(n)^{q_s-q} n^{- \Omega_{k_s}-(p+\Delta p)}+ o(\log(n)^{q_s} n^{- \Omega_{k_s}-p_s^\star}) \right),\end{align}
 where $P$ is a finite set of non-negative reals that contains zero.
Define $\nu_1,\dots,\nu_\omega$ by the $\omega$ lowest levels $\mathcal{E}_s(n) = \Omega_{k_s} + n$ over all singularities $s$ and $n\in \mathbb{Z}_{\geq0}$ (`singularity filling'). Define also $\nu_\star = \min_s\{\Omega_{s} + p_{s}^\star\}$.
 
We can find a basis of mutually anticommuting edge modes
$\hat{\gamma}_L^{(r)}=\sum_{n=0}^\infty \hat{g}_n^{(r)} \gamma_n$ for $1 \leq r\leq \omega $
such $ \hat{g}_n^{(r)}=O(\log(n)^{q_r} n^{-\tilde{\nu}_r})$; here $\tilde{\nu}_r= \min\{\nu_r, \nu_\star\}$; while $q_r$ is equal to $q_s$ for the corresponding singularity.
\end{theorem}

Let us first do an analysis of linear combinations of asymptotic expansions. 
Suppose we have an expansion:
\begin{align}
g_n= \sum_{s=1}^r \sum_{p =0}^\infty \rme^{\rmi n k_s } a_{s,p} n^{- \Omega_{k_s}-p}
\end{align}
Then we have that:
\begin{align}
g_{n-m}= \sum_{s=1}^r \sum_{p =0}^\infty \rme^{\rmi n k_s } \rme^{-\rmi m k_s } a'_{s,p} n^{- \Omega_{k_s}-p \label{eq:gnm}}\end{align}
where $a'_{s,0} = a_{s,0}$ and the other terms can in principle be computed from the expansion of $(n-m)^{- \Omega_{k_s}-p}$. By taking a linear combination $\tilde{g}_n = g_n - A g_{n-m}$ we can cancel the leading term from one (and only one) of the singularities. Indeed, we simply choose $A=\rme^{\rmi m k_s}$ to cancel the leading term of the series about the $s$th singularity. 

Now we show we can cancel terms inductively according to singularity-filling. Suppose we have a set $\{g^{(0)}_n, g^{(1)}_n, \dots ,g^{(m)}_n\}$ constructed from  $\{g_n, g_{n-1}, \dots ,g_{n-m+1}\}$, such that each of the terms $g^{(j)}_n =\sum_{s=1}^r \sum_{p =p_0(s)}^\infty \rme^{\rmi n k_s } a_{s,p} n^{- \Omega_{k_s}-p}$. Here $p_0(s)$ is the `filling' of the singularity $s$. For $j=0$, $p_0(s)=0$ for all $s$, for $j=1$, $p_0(s)=\delta_{s t}$ where $\Omega_{k_t}$ is the minimum of all of the $\Omega_{k_s}$ and so on. Now, we take $g_{n-m}$ to add to our set, this will be of the form \eqref{eq:gnm}. We can then take a linear combination of  $g_{n-m}$ and $g^{(0)}_n$ to cancel the leading term, and get a new expansion $g'_{n-m}$. A linear combination of  $g'_{n-m}$ and $g^{(1)}_n$ will cancel the next leading term (according to the singularity-filling prescription). Continuing in this way we cancel dominant terms until we reach $g^{(m+1)}_n$ which decays faster than $g^{(m)}_n$. Since we always cancel the dominant term we are in accordance with singularity-filling.

Now, suppose the $g_{n-m}$ are the wavefunction coefficients of our linearly independent zero modes $\gamma_L^{(m)}$ as given in Theorem \ref{thm1}. We can take the linear combinations prescribed above and it is clear that we maintain linear independence. For us to have a good basis of edge modes we also need them to mutually anti-commute. This can be achieved by a Gram-Schmidt process \cite{Verresen18}---if we do this in order of fastest decaying to slowest decaying we will preserve the asymptotic decay rates found above. 

Note that it may be the case that when taking linear combinations some $a'_{s,p}$ vanishes accidentally, this simply means we can find even faster decaying modes. We also need to deal with more general asymptotic expansions that have discrete sets of powers as well as logarithmic terms. 

First consider an expansion of the form 
\begin{align}
g_{n}= \sum_{s=1}^r \sum_{p \geq 0} \rme^{\rmi n k_s } a_{s,p} n^{- \Omega_{k_s}-p}
\end{align}
where $p$ is discrete. An example would be the expansion with a single singularity:
\begin{align}
g_{n} = n^{-\Omega} (a_0 + a_1/n +a_2/n^2+\dots) + n^{-\Omega-\alpha} (b_0 +b_1/n +b_2/n^2+\dots),
\end{align}
with $\alpha>0$ (i.e., $P = \{0,\alpha\}$ in Theorem \ref{thm:2}). If we take a linear combination to cancel $a_0$ it will necessarily also cancel $b_0$, so we indeed restrict to the series $\Omega + n$ and ignore $\alpha$, consistent with the claim in Theorem \ref{thm:2}.

Consider now logarithmic terms in the asymptotic expansion. Suppose then:
\begin{align}
g_n= \sum_{s=1}^r \sum_{p\geq 0} \sum_{q =0}^\infty  \rme^{\rmi n k_s } a_{s,p,q} \log(n)^{q_s-q} n^{- \Omega_{k_s}-p}.
\end{align}
Note that now $g_n = \Theta(\log(n)^{q_s} n^{- \Omega_{k_s}})$, where $s$ minimises $\Omega_{k_s}$. Using that $\log(n-m) = \left(\log(n) - \sum_{j=1}^\infty \frac{1}{j}\left(\frac{m}{n}\right)^j\right)$, we have that:
\begin{align}
g_{n-m}= \sum_{s=1}^r \sum_{p\geq0}\sum_{q =0}^\infty \rme^{\rmi n k_s } \rme^{-\rmi m k_s } a'_{s,p,q} \log(n)^{q_s-q} n^{- \Omega_{k_s}-p},
\end{align}
where $a'_{s,0,q}=a_{s,0,q}$. We can then fill singularities inductively as above, where the decay associated to each singularity will be  $O(\log(n)^{q_s} n^{- \Omega_{k_s}-m})$ for $m \in \mathbb{Z}_{\geq 0}$.

To complete the proof, we need to consider the restriction on the sums by $p_s^\star$ and $q_s^\star$. The key point is the error term $o(\log(n)^{q_s} n^{- \Omega_{k_s}-p_s^\star})$, where we do not know the explicit $n$ dependence, and hence the behaviour on taking linear combinations. We can apply singularity filling as described above up to the point this term is no longer subdominant. This leads to the $\nu_\star$ in Theorem \ref{thm:2}.

Having Theorem \ref{thm:2}, we can deduce Theorem \ref{thm2} of the main text using the connection between singularities of $1/f(1/z)$ and $b_-(1/z)^{-1}$. In particular, suppose that $1/f(1/z)$ has an expansion of the form $\sum_{s=1}^r \rme^{\rmi n k_s }n^{-\Omega_{k_s}} \left(\sum_{p\geq 0} a_{s,p} n^{-p} \right)$. Then we can use the discussion in the previous section to see $b_-(1/z)$ has an expansion $\sum_{s=1}^r \rme^{\rmi n k_s }n^{-\Omega_{k_s}} \left(\sum_{p= 0}^{\lfloor \Omega_{\min}\rfloor -2 } a'_{s,p} n^{-p} +O(n^{-\lfloor \Omega_{\min}\rfloor -1})\right)$; thus $\nu^\star =\Omega_{\textrm{min}}+\lfloor \Omega_{\textrm{min}}\rfloor -2$, recovering Theorem \ref{thm2} of the main text.

\subsection{Weierstrass chains}\label{app:Weierstrass}
Here we present an example of a chain where we establish the bulk-boundary correspondence for $t_n$ that decay slower than $1/n$. This is perhaps surprising based on previous literature on long-range chains \cite{Lepori17}, but is straightforward given our condition of absolute-summability. 

Take an integer $b>1$, and a real number $\beta>0$. The corresponding Weierstrass chain has couplings 
\begin{align}
t_n = \begin{cases}
\mu \qquad & n =0\\
b^{-m \beta} \qquad &n = \pm b^m \qquad m \in \mathbb{N}\\
0 \qquad &\textrm{otherwise}.
\end{cases}
\end{align}
For $\mu >1/ (b^\beta-1)$, the corresponding $f(z=\rme^{\rmi k}) = \mu+ \sum_{n\geq 1} b^{-n \beta} \cos(b^n k)$ is gapped and has winding number zero. The series converges absolutely \cite{Zygmund02,Grafakos08}, and one can see that $t_n \leq C/(1+\vert n\rvert)^\beta$; i.e., the $t_n$ are $\beta$-decaying. Hence, we have examples of $\beta$-decaying chains for any $\beta>0$. We can then use Theorem \ref{thm1} to find edge modes for the shifted Weierstrass chains $f(z) \rightarrow z^\omega f(z)$. Note that for $0<\beta<1$, $f(z)$ is nowhere differentiable \cite{Zygmund02}. So long as we maintain the gap, we can add couplings $t_n$ corresponding to another absolutely-summable chain; this means we can find further $(\beta>0)$-decaying chains that are not restricted to the special case studied here where many $t_n=0$. 
\subsection{Short-range chains}\label{app:shortrange}
Here we connect Theorem \ref{thm1} to the known results in the finite-range case \cite{Motrunich01,DeGottardi13,Verresen18}. Indeed, in this case, the results reduce to those given in Ref.~\cite{Verresen18}. 

If the $t_n$ are non-zero for only a finite range, then we have that:
\begin{align}
f(z) = \rho \frac{1}{z^{N_p}} \prod_{j=1}^{N_z} \left(z-z_j\right)\prod_{k=1}^{N_Z} \left(z-Z_k\right) \qquad\qquad \rho \in \mathbb{R}\setminus\{0\}. 
\end{align}
The $z_j$ are inside the unit circle, and $Z_k$ are outside the unit circle. We can read off $\omega=N_z-N_p $, and for $\omega>0$ we have that there are $\omega$ edge modes and there is a basis where the localisation lengths are set by the $\omega$ zeros closest to the unit circle. A proof is given in Ref.~\cite{Verresen18}.

We can get the same result using our Theorem \ref{thm1}. First, up to a factor $\rme^{V_0}$ that we fix by rescaling the Hamiltonian, we have:
\begin{align}
f(z) = z^{\omega} \underbrace{\prod_{j=1}^{N_z} \left(1-z_j/z\right)}_{b_-(z)}\underbrace{\prod_{k=1}^{N_Z} \left(1-z/Z_k\right)}_{b_+(z)}.
\end{align}
For $\omega>0$ we then use Theorem \ref{thm1} to identify $\omega$ edge modes with wavefunctions given by the Fourier coefficients:
\begin{align}
g_n^{(m)} =\frac{1}{2\pi\rmi}\int_{S_1} \frac{1}{\prod_{j=1}^{N_z} (1-zz_j) }z^{m-n-1} \rmd z = \sum_{j=1}^{N_z} a_j^{\vphantom{n-m}} z_j^{n-m}, \label{eq:shortrangeedge}
\end{align}
for $n$ sufficiently large and where $a_j^{-1} = \prod_{k\neq j}(1-z_k/z_j)$. As in the proof of Theorem \ref{thm2} (and in corresponding analysis in \cite{Verresen18}) we can then take appropriate linear combinations to get the claimed localisation lengths. 

An analogous discussion holds for $\omega<0$ and zeros outside the unit circle appearing in $b_+(z)$. Moreover, we can use the analysis of gapless models given in the main text (see also below) to see that short-range gapless models have edge modes with localisation lengths determined by zeros of $f(z)$.

One may consider long-range chains as a limiting case of short-range chains, where the interaction range tends to infinity. Then, the degree of the pole and/or the number of zeros on the unit circle increases without bound. The results in this paper apply to fixed Hamiltonians. This means that short-range chains (even with arbitrarily large but finite range) have edge-modes with exponentially-decaying wavefunctions for sufficiently large site index. On the other hand, long-range chains, even with very weak long-range couplings (e.g., $\alpha$-decaying models with arbitrarily large $\alpha$), will typically have an algebraically-decaying wavefunction for sufficiently large site-index. Physically we expect that these cases should behave similarly; the starkly different behaviours for a fixed Hamiltonian are a consequence of the particular sequence of limits that we are working in. For short-range chains with a large finite range, the edge mode will have a wavefunction corresponding to \eqref{eq:shortrangeedge}, and for a certain values of $n$ (depending on the range of the Hamiltonian) this will approximate the algebraic decay of a long-range chain. Hence, by considering a sequence of finite-range Hamiltonians converging to a long-range model, we expect to see agreement. This is comparable to the approximation of the ground-states of critical spin chains using a sequence of matrix-product states of increasing bond dimension \cite{Andersson99}. 
\section{Calculations for first example}\label{app:example}
\subsection{Set up}
We will consider  $f(z) = z^\omega \textrm{Li}_\beta(z) \textrm{Li}_\gamma(1/z)$, the case studied in the main text follows by putting $\beta = \gamma=\alpha$. This $f(z)$ corresponds to Hamiltonian couplings of the form \begin{align}
t_n = \begin{cases}\sum_{k=n+1}^\infty \frac{1}{k^\beta (k-n)^{\gamma}} \qquad &n\geq 0\\
\sum_{k=|n|+1}^\infty \frac{1}{(k-|n|)^\beta k^{\gamma}} \qquad &n<0.
\end{cases}
\end{align}
We assume that $\beta>1$, $\gamma>1$, and, in order to compute the edge mode asymptotics, that they are non-integer.
\subsection{Decay of couplings}
We first show that $t_n$ are $\alpha_0$-decaying with $\alpha_0 = \min(\beta,\gamma)$ as follows. First, for $n\geq 0$: 
\begin{align}
t_n = \sum_{k=n+1}^\infty \frac{1}{k^\beta (k-n)^{\gamma}}=\frac{1}{(n+1)^\beta} \sum_{k=n+1}^\infty \left(\frac{n+1}{k}\right)^\beta \frac{1}{ (k-n)^{\gamma}}
\leq \frac{1}{(1+n)^\beta} \sum_{k=1+n}^\infty \frac{1}{ (k-n)^{\gamma}} = \frac{\zeta(\gamma)}{(n+1)^\beta}. \end{align}
The analogous calculation for $n<0$ gives $t_n \leq \zeta(\beta)/(1+\lvert n \rvert)^\gamma$. Hence, we have an upper bound for all $n$ of $t_n \leq \zeta(\alpha_0)/(1+\lvert n \rvert)^{\alpha_0}$. Since we also have $t_n \geq (1+n)^{-\beta}$ for $n\geq 0$ and $t_n \geq (1+|n|)^{-\gamma}$ for $n<0$ we have for large and positive [negative] $n$, $t_n = \Theta (n^{-\beta})$ [$t_n =\Theta( n^{-\gamma})$].

\subsection{Asymptotics of edge mode}
Now we calculate the asymptotic form of the edge mode wavefunction.
\begin{align}
g_n = \frac{1}{2\pi \rmi}\int_{S^1} \frac{z^{-n}}{\textrm{Li}_\gamma(z)}\rmd z . \label{eq:fouriercontour}
\end{align}
First, $\textrm{Li}_\gamma(z)$ has a branch point singularity at $z=1$, and we analytically continue to the plane with a branch cut $z\in [1,\infty)$. Note that $\textrm{Li}_\gamma(z)$ is non-zero for $z\neq0$, so  $\textrm{Li}_\gamma(z)^{-1}$ has no poles for $z\neq0$ \cite{Wood92}. Using the integral representation of $\textrm{Li}_\gamma(z)$ \cite[Eq.~25.12.11]{NIST:DLMF} we have that  $\textrm{Li}_\gamma(\rme^x-\rmi\eps) = \overline{\textrm{Li}_\gamma(\rme^x+\rmi\eps)}$ for $x\geq 0$ and as $\eps \rightarrow 0$.
Now, assuming $\gamma$ is not an integer, we have the expansion \cite[Eq.~25.12.12]{NIST:DLMF}:
\begin{align}
\lim_{\eps\rightarrow 0} \textrm{Li}_\gamma(\rme^x-\rmi \eps) = \Gamma(1-\gamma) \rme^{\rmi \pi (\gamma-1)} x^{\gamma-1} + \sum_{n=0}^\infty \zeta(\gamma-n)\frac{ x^n}{n!}\qquad |x|<2\pi. \label{eq:asymptoticpolylog}
\end{align}

Deforming the contour in \eqref{eq:fouriercontour} out to infinity leaves us with a branch cut contribution:
\begin{align}
g_n &= \lim_{\eps\rightarrow0}\frac{1}{2\pi\rmi}\Bigg( \int_0^\infty \frac{\rme^{- n x}}{\textrm{Li}_\gamma(\rme^x+\rmi \eps)}  \rmd x - \int_0^\infty \frac{\rme^{- n x}}{\textrm{Li}_\gamma(\rme^x-\rmi \eps)} \rmd x \Bigg)
\\&=  \lim_{\eps\rightarrow0}\frac{1}{2\pi\rmi}\Bigg( \int_0^\infty \frac{\rme^{- n x}}{\overline{\textrm{Li}_\gamma(\rme^x-\rmi \eps)}}  \rmd x - \int_0^\infty \frac{\rme^{- n x}}{{\textrm{Li}_\gamma(\rme^x-\rmi \eps)}} \rmd x \Bigg)
\\&= \frac{1}{\pi} \lim_{\eps\rightarrow0} \Im \int_0^\infty \frac{ \textrm{Li}_\gamma(\rme^x-\rmi \eps)\rme^{- n x}}{\lvert \textrm{Li}_\gamma(\rme^x-\rmi \eps)\rvert^2}  \rmd x .
\end{align}
Then, we can insert the expansion \eqref{eq:asymptoticpolylog}, valid in the region near the end of the branch cut, and conclude that:
\begin{align}
g_n&= -\frac{1}{\pi}  \frac{\Gamma(1-\gamma) \sin(\pi(1-\gamma))}{n^\gamma \zeta(\gamma)^2}\int_0^\infty x^{\gamma-1}\rme^{- x} (1+ o(1) ) \rmd x \\
&=  - \frac{1}{n^\gamma \zeta(\gamma)^2} (1+ o(1) ). 
\end{align}
To get the next algebraic correction we simply take the next term in the expansion of $\lvert \textrm{Li}_\gamma(\rme^{x/n}-\rmi \eps)\rvert^{-2}$ using \eqref{eq:asymptoticpolylog}.

\section{The long-range Kitaev chain}\label{app:Kitaev}

\subsection{Edge mode wavefunctions}\label{app:wavefunction}
Recall that the long-range Kitaev chain is of the form:
\begin{align}
    f_\textrm{LRK}(z)    &=\mu + J\left(\mathrm{Li}_\alpha(z)+\mathrm{Li}_\alpha(1/z)\right)+ \Delta\left(\mathrm{Li}_\beta(z)-\mathrm{Li}_\beta(1/z)\right).\label{eq:kitaevapp}
\end{align}
We assume here that the parameters are chosen so that $f_\textrm{LRK}(z)$ has $\omega=1$. 

Aside from certain special cases, we do not have a closed-form for the Wiener-Hopf decomposition (see the next subsection for an example of such a special case), as would be needed to find the exact edge-mode as given in Theorem \ref{thm1}. 
However, as discussed above, the asymptotic Fourier coefficents of  $1/f_\textrm{LRK}(1/z)$ and the corresponding $1/b_-(1/z)$ have the same singularities.
Hence, for the topological phase, we can find the \emph{asymptotic decay} of the wavefunction by calculating large Fourier coefficients, $G_n$, of $1/f_\textrm{LRK}(1/z)$ (one could also use the statement of Theorem \ref{thm2} directly).

The calculation of these large Fourier coefficients is similar to that given in the previous section and goes as follows. By definition,
\begin{align}
G_n = \frac{1}{2\pi \rmi}\int_{S^1} \frac{z^{-n-1}}{\mu + J\left(\mathrm{Li}_\alpha(z)+\mathrm{Li}_\alpha(1/z)\right)- \Delta\left(\mathrm{Li}_\beta(z)-\mathrm{Li}_\beta(1/z)\right)}\rmd z . \label{eq:1/f}
\end{align}
The integrand is analytic in the same cut-plane as $f_{\textrm{LRK}}(z)$, excluding isolated poles at $z=0$ and $z=1/z_j$ where $z_j$ are the zeros of $f_{\textrm{LRK}}$. We deform the contour out to infinity, and the contour gets snagged on the branch cut and at the poles outside the unit circle. The contribution from these poles will decay exponentially as $\Theta(z_j^n n^{k-1})$ for some $k\in \mathbb{N}$, corresponding to the degree of the pole.

Using that $\mathrm{Li}_\alpha(\rme^{-x})$ is real for $x>0$ \cite[Eq.~25.12.11]{NIST:DLMF}, and continuous across the branch cut, the integral along the branch cut is:
\begin{align}
\lim_{\eps\rightarrow 0}  \frac{1}{\pi } \int_{0}^\infty\rme^{-(n+1)x} \frac{\Im [J~\mathrm{Li}_\alpha(\rme^x -\rmi \eps)-\Delta\mathrm{Li}_\beta(\rme^x -\rmi \eps)]}{\big\lvert\mu + J\left(\mathrm{Li}_\alpha(\rme^x -\rmi \eps)+\mathrm{Li}_\alpha(\rme^{-x})\right)- \Delta\left(\mathrm{Li}_\beta(\rme^x -\rmi \eps)-\mathrm{Li}_\beta(\rme^{-x})\right)\big\rvert^2}\rmd x.\label{eq:LRKedge}\end{align} 
Using the expansion \eqref{eq:asymptoticpolylog}, and that all other contributions are exponentially decaying, we have that:
\begin{align}
G_n = - \frac{1}{ (\mu+2J\zeta(\alpha))^2}\Big(J n^{-\alpha} (1+o(1)) -\Delta n^{-\beta}(1+o(1))  \Big) + O(z_0^n n^{k-1}) \end{align}
where $z_0$ is the zero of $f_\textrm{LRK}$ inside of and closest to the unit circle, $k\in \mathbb{N}$, and to evaluate the branch cut contribution we also use the expansion  \cite[Eq.~25.12.12]{NIST:DLMF}:
\begin{align}
\textrm{Li}_\gamma(\rme^{-x}) = \Gamma(1-\gamma)  x^{\gamma-1} + \sum_{n=0}^\infty \zeta(\gamma-n)\frac{ (-x)^n}{n!}\qquad |x|<2\pi. \label{eq:asymptoticpolylog2}
\end{align}
We hence see that in general, the edge mode decays as $n^{-\min(\alpha,\beta)}$.  For the case where $J=\Delta$ and $\alpha=\beta$, the branch cut integral vanishes, and we have exponentially localised modes, with localisation length $\xi^{-1} = -\log(|z_0|)$. We can take further terms in the expansion of the denominator in \eqref{eq:LRKedge} to derive subdominant terms in the asymptotic expansion, noting that they decay as inverse powers of $n$ and so we have a nice expansion for these Fourier coefficients.

 With different justifications, two closely related integrals to \eqref{eq:1/f} were analysed in Refs.~\cite{Jaeger20,Francica22}, leading to the same conclusion: a single edge mode decays as $n^{-\min(\alpha,\beta)}$. This result was moreover in agreement with previous numerical results \cite{Vodola14,Vodola15,Alecce17}.

Our method not only gives an analytic approach to finding the asymptotic decay of the single edge mode in this model, it also allows us to predict the decay of the edge modes for higher winding numbers, as discussed in the main text (although we note that finite-size effects can hybridise the edge modes and so the energy eigenbasis in a finite chain may not have the same form as the basis given in Theorem \ref{thm2}).

\subsection{Finite-size splittings}\label{app:example2}
Let us consider a case of the long-range Kitaev chain \eqref{eq:kitaevapp} when $\alpha=\beta$ and $J=\Delta$, and consider $\lvert a\rvert =\lvert 2J/\mu\rvert<\zeta(\alpha)^{-1}$, we are then in the trivial phase. We will use the rigorous methods explained in the next section (Rigorous Underpinnings for Conjecture \ref{conjecture}) to calculate the relevant Toeplitz determinant that we believe gives us the edge mode splittings.
After an overall renormalisation:
\begin{align}
f_0(z)=  1 + a\mathrm{Li}_\alpha (z) = b_+(z) \label{eq:example2}.
\end{align}
In this case, we have that $1/f_0(z)$ has no negative Fourier coefficients. We can then use Theorem \ref{thm:BoettcherSilbermann} (given below) to see that the splitting is exactly zero if we take $f(z)\rightarrow z^\omega f_0(z)$ for $\omega>0$. This is trivial to observe at the Hamiltonian level, we have decoupled Majorana modes for this choice of $f(z)$. More interesting is that for $\omega<0$ the same theorem leads us to
predict edge mode splittings as in Proposition \ref{prop:asymptotics}.
In particular $l(z) = \frac{1}{1+a\mathrm{Li}_\alpha (z)}$, and we can calculate (as in the previous section):
\begin{align}
l_N = -\frac{a}{(1+a \zeta(\alpha))^2} N^{-\alpha} (1 + \dots)     
\end{align}
where the further terms in the expansion come from the same branch-cut integral and do not oscillate. Note that the asymptotic result $D_N[z^\omega f_0(z)] = \Theta( N^{-\lvert\omega\rvert(\lvert\omega\rvert-1) -\lvert\omega\rvert\alpha})$ is rigorous (see the following section, and in particular Theorem \ref{thm:BoettcherSilbermann}). 

Thus we expect based on the singularity-filling picture that the model $f(z) = z^\omega (1+a\mathrm{Li}_\alpha(z))$ for $\omega<0$ has edge modes with single-particle energies
$L^{-\alpha}, L^{-\alpha - 2},\dots L^{-\alpha -2(\lvert\omega\rvert -1)}$.

\subsection{Further numerical results:}\label{app:numerics}
\begin{figure}
    \centering
    \includegraphics[]{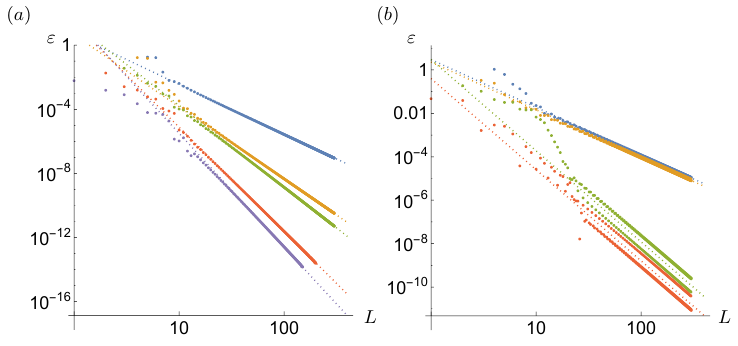}
    \caption{(a), (b) Finite-size splittings for the model \eqref{eq:appf0} for $\omega =5$ and $\omega=-4$ respectively. We see singularity-filling (dashed lines) correctly predicting the splittings. Note that (b) features oscillatory behaviour corresponding to the complex singularities.}
    \label{fig:2}
\end{figure}

 A generalisation of the long-range Kitaev chain, defined by \begin{align}f(z) = z^{\omega} \underbrace{\left( 2+ \mathrm{Li}_{2.2}(\rme^{\rmi\pi/3}z)+ \mathrm{Li}_{2.2}(\rme^{-\rmi\pi/3}z)+\mathrm{Li}_{4.5}(1/z)+\mathrm{Li}_{3.1}(-1/z) \right)}_{f_0(z)} \label{eq:appf0}\end{align} is considered in the main text. The singularities are depicted in Fig.~1, and the singularity-filling picture for finite-size splitting is confirmed there for $\omega=4$. For $\omega=5$ we expect the same behaviour for the four splittings $\eps_1,\dots\eps_4$ (this is our inductive assumption in the main text), and an additional $\eps_5=\Theta(L^{-7.1})$. 

For $\omega<0$ the branch-cuts inside the circle do not contribute, and the branch-cuts outside the circle are relevant. For $\omega=-4$ we hence expect $\eps_1, \eps_2 =\Theta( L^{-2.2})$ and $\eps_3, \eps_4 =\Theta( L^{-4.2})$. We also have singularities at momentum $\theta=\pm \pi/3$, so we expect oscillations $\rme^{\rmi L m \pi/3}$ for integer $m$ (see also the next section where we observe that the oscillations appear in the Toeplitz determinant). All of these expectations are confirmed in Fig.~\ref{fig:2}(a) and (b).
\section{Rigorous underpinnings for Conjecture \ref{conjecture}}\label{app:filling}
\subsection{An upper bound on the finite-size splitting}\label{app:split}
Here we prove the following proposition regarding the finite-size splitting.
\begin{proposition}
Suppose that the Hamiltonian $H_{\mathrm{BDI}}$ is $(\alpha>2)$-decaying, and has $\omega \geq1$. Consider the fastest-decaying edge mode on a half-infinite system $\gamma_\star = \sum_{n=0}^\infty b_n \gamma_n $, with wavefunction asymptotics $b_n = \mathrm{const}\times n^{-\nu} (1+o(1))$. Then for the corresponding system of size $(L+1)$ with open boundaries, the finite-size splitting, $\eps$, is $O(L^{-\nu})$.
\end{proposition}
This is a weak form of Conjecture \ref{conjecture}, for the case $\omega =1$ (and for $\omega =-1$ with the usual replacements), since we conjecture the splitting is $\Theta(L^{-\nu})$, and the proposition is restricted to $(\alpha>2)$-decay. The result applies also for $\lvert \omega \rvert >1$, giving an even weaker form of Conjecture \ref{conjecture} in such cases, since we conjecture the splitting is $\Theta(L^{-\nu'})$ where $\nu'\geq \nu$, where the inequality is strict if any singularity is filled more than once.

To prove this result, we split the half-infinite edge mode into a mode supported on a region $A$, consisting of the sites 0 up to $L$, and the region $B$ consisting of the remaining sites.
\begin{align}
\gamma_\star = \frac{1}{\mathcal{Z}_L}\sum_{n=0}^\infty b_n \gamma_n =\underbrace{ \frac{1}{\mathcal{Z}_L}\sum_{n=0}^L b_n \gamma_n}_{\gamma_A} + \underbrace{ \frac{1}{\mathcal{Z}_L}\sum_{n=L+1}^\infty b_n \gamma_n}_{\gamma_B} .
\end{align}
We choose the normalisation $\mathcal{Z}_L$ so that $\gamma_A^2=1$. In the large $L$ limit, $\mathcal{Z}_L$ tends to a constant; we suppress this from the notation below.
Similarly we split the Hamiltonian $H_{\mathrm{BDI}}=\rmi \sum_{n \geq 0,m\geq 0} t_{m-n}  \tilde\gamma_n \gamma_{m} $ on a half-infinite chain as:
\begin{align}
H_{\mathrm{BDI}} = \underbrace{\rmi \sum_{0\leq n,m\leq L } t_{m-n}  \tilde\gamma_n \gamma_{m}}_{H_{AA}}+  \underbrace{\rmi \sum_{m=L+1}^\infty \sum_{ n=0 }^L t_{m-n}  \tilde\gamma_n \gamma_{m}}_{H_{AB}}+  \underbrace{\rmi \sum_{n=L+1}^\infty \sum_{ m=0 }^L t_{m-n}  \tilde\gamma_n \gamma_{m}}_{H_{BA}}+  \underbrace{\rmi \sum_{ m,n>L } t_{m-n}  \tilde\gamma_n \gamma_{m}}_{H_{BB}}.
\end{align}
Notice that $H_{AA}$ is the Hamiltonian for a finite chain of size $L+1$ with open boundaries.
Let $\ket{\psi}$ be the ground state for the half-infinite chain with Hamiltonian $H_{AA} +  \sum_{n=L+1}^\infty c_n^\dagger c_n^{\vphantom\dagger}$. The state $\gamma_A \ket{\psi}$ is orthogonal to the ground state, and we can consider the variational energy in this parity sector relative to the ground state: 
\begin{align}
\tilde\eps &=\bra{\psi}[\gamma_A , H_{AA}] \gamma_A \ket{\psi}=\bra{\psi}[\gamma_\star, H_{AA}] \gamma_A \ket{\psi}\nonumber\\
&=\bra{\psi}[\gamma_\star, H-H_{AB}-H_{BA}-H_{BB}] \gamma_A \ket{\psi}.
\end{align}
Since $\gamma_A$ has support only on $\gamma_n$ for $n\in A$, and $\ket{\psi}$ has no correlations between $A$ and $B$, this reduces to: 
\begin{align}
\tilde\eps &=-\bra{\psi}[\gamma_\star, H_{AB}] \gamma_A \ket{\psi}.
\end{align}
Now:
\begin{align}
[\gamma_\star, H_{AB}]& = \rmi \sum_{k=0}^\infty  \sum_{m=0}^\infty \sum_{n=0}^L b_k t_{m+L+1-n}[ \gamma_k , \tilde{\gamma}_{n}\gamma_{m+L+1}]\\
&= -2 \rmi  \sum_{m=0}^\infty \sum_{n=0}^L b_{m+L+1} t_{m+L+1-n} \tilde{\gamma}_{n}
\end{align}
Hence,
\begin{align}
\tilde\eps &=2 \rmi \sum_{k=0}^L  \sum_{m=0}^\infty \sum_{n=0}^L  b_k b_{m+L+1}t_{m+L+1-n} \bra{\psi} \tilde{\gamma}_{n}     \gamma_k \ket{\psi}.
\end{align}
Two-point correlations are upper bounded by $\lvert \rmi \langle \tilde\gamma_n \gamma_m \rangle \rvert \leq 1$.
Hence:
\begin{align}
\lvert \tilde\eps \rvert &\leq 2  \sum_{k=0}^L  \sum_{m=0}^\infty \sum_{n=0}^L  \lvert b_k\rvert  \lvert b_{m+L+1}\rvert\lvert t_{m+L+1-n} \rvert= 2\left( \sum_{k=0}^L \lvert b_k \rvert \right)\left( \sum_{m=0}^\infty  \sum_{n=0}^L  \lvert b_{m+L+1}\rvert\lvert t_{m+L+1-n}\rvert \right) .
\end{align}
Since we have that the $b_k$ form an absolutely convergent Fourier series (of the function $1/b_-(1/z)$), the first term is upper bounded by a constant. I.e.,
\begin{align}
\tilde\eps &\leq \textrm{const}\times \left( \sum_{m=0}^\infty  \sum_{n=0}^L  \lvert b_{m+L+1}\rvert \lvert t_{m+L+1-n}\rvert \right)= \textrm{const}\times  \left( \sum_{m=0}^\infty  \sum_{n=0}^L  \lvert b_{m+1+L}\rvert\lvert t_{m+1+n}\rvert\right) .
\end{align}
Now, suppose that for large $n$, $\lvert b_n \rvert \propto n^{-\nu}(1+o(1))$, then:
\begin{align}
\tilde\eps &\leq \textrm{const} \times L^{-\nu}(1+o(1))\left( \sum_{m=0}^\infty \sum_{n=0}^L \lvert t_{m+1+n}\rvert\right) \leq  \textrm{const} \times L^{-\nu}(1+o(1))\left(\lim_{M\rightarrow \infty}  \sum_{n=1}^{2M+1} n \lvert t_{n} \rvert +o(1) \right) .
\end{align}
Since we assume that the $t_n$ are $(\alpha>2)$-decaying, this last expression is summable, and so we have that $\tilde\eps = O(  L^{-\nu})$. Hence by the variational method, the finite-size splitting $\eps= O(L^{-\nu})$.
\subsection{Notation}
Denote the $L\times L$ Toeplitz determinant generated by $t(z)$ by $D_L[t(z)]$. In the main text we make the connection between the decay of the product $\prod_{j=1}^L(-\eps_j^2)$ and the edge mode splitting. The first thing to note is that the single-particle Hamiltonian for our chain is a block Toeplitz matrix generated by:
\begin{align}
\Phi(z) = \left(\begin{array}{cc}0 & f(z) \\-f(1/z) & 0\end{array}\right).
\end{align}
Then $\prod_{j=1}^L(-\eps_j^2)=D_L[\Phi(z)] = (-1)^L D_L[f(z)]D_L[f(1/z)]=(-1)^L D_L[f(z)]^2 $. Hence, we have that:
\begin{align}
\prod_{j=1}^L \eps_j = \lvert D_L[f(z)]\rvert = \lvert D_L[z^\omega f_0(z)] \rvert.
\end{align}
By calculating $D_L[z^\omega f_0(z)]$ we can estimate the edge mode decays with the assumptions made in the main text. Note that it if we rescale $f(z) \rightarrow c f(z)$, this product will be rescaled by $c^L$. There is a natural overall normalisation that we use implictly throughout (in particular we fix $V_0=0$ below).

Notational remark: for clarity in various formulae, in this section we use $\theta\in[0,2\pi)$ to denote momenta on the unit circle, rather than $k$, and the finite number of singularities on the unit circle are denoted by $\theta_s$ rather than $k_s$.
\subsection{Toeplitz determinants}
\subsubsection{Definitions.}
Suppose that $f(z)$ corresponds to a gapped, $\alpha$-decaying Hamiltonian with $\alpha>1$. Then we can write $f(z) = z^\omega f_0(z)$, where $f_0(z)\neq0$ on the unit circle. There exists a $V(z)$ that is a continuous logarithm of $f_0(z)$, i.e., $f_0(z) =\rme^{V(z)}$. 

Then by the Wiener-L\'evy theorem, we have that the following Fourier series converges absolutely:
\begin{align}
V(z) = \sum_{n=-\infty}^\infty V_n z^n.
\end{align}

We can then define:
\begin{align}
b_+(z) &= \rme^{\sum_{n=1}^\infty V_n z^n}\\
b_-(z) &= \rme^{\sum_{n=1}^\infty V_{-n} z^{-n}}\end{align}
so that\begin{align}
f(z) &= b_+(z) \rme^{V_0}b_{-}(z).
\end{align}
We fix $V_0=0$ by a rescaling. Note that $b_+(0)=b_{-}(\infty)=1$, $b_+(z)$ is analytic inside the disk $\lvert z \rvert< 1$, and $b_-(z)$ is analytic for $\lvert z \rvert> 1$. 

We can then define the functions:
\begin{align}
l(z) = \frac{b_-(z)}{b_+(z)} \qquad\qquad m(z) =\frac{b_+(1/z)}{b_-(1/z)};
\end{align}
these functions also have an absolutely convergent Fourier expansion. 
\subsubsection{Szeg\H{o}'s theorem}
Under the previous assumptions, we can evaluate the asymptotics of $D_N[f_0(z)]$  as:
\begin{align}
D_N[f_0(z)] = \rme^{V_0 N} \rme^{\sum_{k=1}^\infty k V_k V_{-k}}(1+o(1)),
\end{align}
note that our conditions guarantee that $\rme^{\sum_{k=1}^\infty k V_k V_{-k}}\neq0$, as claimed in the main text.

\subsubsection{Some results on shifted determinants}
Our method for calculating the edge mode splittings requires the asymptotics of $D_N[z^\omega f_0(z)]$. These determinants are related to the functions $l(z)$ and $m(z)$; analysis and a general result can be found in Ref.~\cite{Hartwig69}. Roughly speaking, for $\omega=1$ the edge mode splitting decays like $m_N$, while for $\omega=2$ the product of edge mode splittings behaves like $m_N^2- m_{N-1}m_{N+1}$, this has some cancellations and behaves like a discrete derivative, leading to the singularity-filling picture. In general this statement holds only up to some error terms that depend on the analytic properties of $f_0(z)$. 

To give a sharper statement, we will use the following theorem from Ref.~\cite{Boettcher06b}:
\begin{theorem}[Fisher, Hartwig, Silbermann et al.]\label{thm:HartwigFisher}
Suppose that $f_0(z)$ 
belongs to $C^\beta$ for $\beta>1/2$ and $\beta \notin \mathbb{Z}$. 
\begin{align}
D_N[z^\omega f_0(z)] = (-1)^{N\omega}D_{N+\lvert\omega\rvert}[f_0(z)]\left( \det(M(N))+O(N^{-3\beta})\right)\left(1+O(N^{1-2\beta})\right)\label{FHS}
\end{align}
where $M(N)$ is an $\lvert \omega\rvert\times\lvert\omega\rvert$ matrix with matrix elements:
\begin{align}
M(N)_{j-k} = \begin{cases} 
l_{N+j-k} \qquad &\omega<0 \\
m_{N+j-k}\qquad &\omega> 0.
\end{cases}
\end{align}
\end{theorem}
Functions in the class $C^\beta$ have $n=\lfloor \beta \rfloor$ continuous derivatives and the $n$th derivative satisfies a H\"{o}lder condition 
\begin{align} \lvert f_0^{(n)}(\rme^{\rmi \theta_1})-f_0^{(n)}(\rme^{\rmi \theta_2})\rvert \leq M_{f_0} \lvert \rme^{\rmi \theta_1}-\rme^{\rmi \theta_2}\rvert^{\beta_0}~~ \forall \theta_1,\theta_2 \in [0,2\pi), \label{eq:Hoelder}\end{align}
where $\beta = n +\beta_0$. 
As we will see below, the asymptotics of $\det(M(N))$ will decay faster as $\lvert \omega \rvert$ increases. Hence, Theorem \ref{thm:HartwigFisher} is limited when looking at large values of $\omega$, where $\det(M(N))$ can be of the same order as the unspecified error term (this motivates the $\omega_\textrm{max}$ in Conjecture \ref{conjecture}). We can evaluate the asymptotics of $\det(M(N))$ using Proposition \ref{prop:asymptotics} (corresponding to singularity-filling, see below) if we have an appropriate asymptotic expansion for $l(z)$ or $m(z)$. 

For the models considered in the main text, we have that $1/f(1/z)\in C^{\beta-1}$ for $\beta =\Omega_{\textrm{min}}-\eps$ where $\eps>0$. Using Proposition \ref{prop:asymptotics} below, we have that the decay of $\det(M(N))$ is at most $N^{-\gamma}$ where $\omega\Omega_{\textrm{min}}\leq \gamma\leq\omega \Omega_{\textrm{min}}+\omega(\omega-1)$. 
We are then justified in using the singularity filling picture \footnote{More carefully: Proposition \ref{prop:asymptotics} and Theorem \ref{thm:HartwigFisher} give singularity-filling for Toeplitz determinants up to $\omega=3$. If we additionally assume that the spectrum is such that the edge mode energies correspond to the Toeplitz determinant according to the inductive method outlined in the main text, then we have singularity filling for the splittings (Conjecture \ref{conjecture}).} as long as $\gamma<3 (\beta -1)< 3(\Omega_{\min}-1)$. This inequality is violated for $\omega =3$, while it is satisfied for $\omega =2$ as long as $\Omega_{\textrm{min}}>5$. (Note: we also need a nice expansion for $m_n$ (or $l_n$), and as proved above we can use the nice expansion for $1/f(1/z)$ to infer this up to the first subleading term whenever $\Omega_{\textrm{min}} >3$.)

Note that this is a conservative estimate for the applicability of Conjecture \ref{conjecture}, since it is based on the possibility of the subleading term in Theorem \ref{thm:HartwigFisher} becoming relevant. Numerics such as Figure~1 in the main text and Figure \ref{fig:2} indicate that, in those models, the singularity filling continues to apply for higher winding numbers.

The following result from \cite{Boettcher06} is useful in certain special cases, including when we have $f(z)$ depending only on $z$:
\begin{theorem}[Boettcher and Silbermann]\label{thm:BoettcherSilbermann}
Suppose $f_0(z)$ satisfies the conditions  $f_0(z)\neq0$ on the unit circle, has winding number zero and has absolutely convergent Fourier series \cite{Devinatz,Widom60}. Furthermore, suppose that the $n$th Fourier coefficient of $1/f_0(z)$ is zero for $n<-n_0\leq 0$. Then for $N\geq n_0$:
\begin{align}
D_N[z^\omega f_0(z)] &= (-1)^{N\omega}D_{N+\lvert\omega\rvert}[f_0(z)] D_{\lvert \omega \rvert}[z^{-N} l(z)] \qquad& \omega<0 \nonumber\\
D_N[z^\omega f_0(z)] &= 0 \qquad &\omega>0
\end{align}
Now suppose that the $n$th Fourier coefficient of $1/f_0(z)$ is zero for $n>n_0\geq 0$. Then for $N\geq n_0$:
\begin{align}
D_N[z^\omega f_0(z)] &= 0 \qquad &\omega<0 \nonumber\\
D_N[z^\omega f_0(z)] &= (-1)^{N\omega}D_{N+\lvert\omega\rvert}[f_0(z)] D_{\lvert \omega \rvert}[z^{-N} m(z)] \qquad &\omega>0.
\end{align}
\end{theorem}
This exact formula  on the right-hand-side means we can analyse the asymptotics without limits on the winding. We do this analysis in the next subsection.
The conclusion is that:
\begin{proposition}\label{prop:asymptotics}
Suppose that $l(z)$ has an asymptotic expansion of the form:
\begin{align}l_N &= \sum_{n=1}^m \sum_{p \geq \Omega_{\theta_n}} \rme^{\rmi N\theta_n } a_{n,p} N^{-p} \qquad a_{n,\Omega_{\theta_n}} = G_n \label{eq:ln}\end{align}
then for $\omega <0$:
\begin{align}
D_{\lvert \omega \rvert}[z^{-N} l(z)]  &=  N^{-e}\sum_{k_1 \dots k_m}((-1)^{\sum_{n=1}^mk_n(k_n-1)/2}c_{k_1,\dots k_m}\prod_{n=1}^m\left(G_n^{k_n}\rme^{\rmi N \theta_n k_n}\right) ) (1+o(1))
\end{align}
where $c_{k_1,\dots k_m}>0$,\begin{align}
e=\min\{ \sum_{n}\left( k_n(k_n-1)+\Omega_{\theta_n} k_n \right)\quad \mathrm{for~}  k_n \in \mathbb{Z}_+, \sum k_n =\lvert \omega \rvert\},\label{eq:e}
\end{align}
and the sum over $k_1,\dots k_m$ is over all choices where this minimum is achieved. If the minimum is unique then we are guaranteed that this is the dominant term for all $N$, otherwise the sum may contain cancellations.
\end{proposition}
The proof relies on a truncation of \eqref{eq:ln}, so we can also consider cases where we have a nice expansion of $l_n$ up to some power. Note that an identical proposition can be written for $\omega>0$ and where the parameters correspond to the asymptotic expansion of $m_N$ (note that these will in general be different to the parameters corresponding to $l_N$). Using this proposition, we can evaluate the asymptotics of Theorem \ref{thm:BoettcherSilbermann}. The decay of this determinant $N^{-e}$ is consistent with the singularity-filling picture [indeed, one can think of the formula \eqref{eq:e} for $e$ as singularity-filling, it is in this sense that we say singularity-filling is rigorous for certain Toeplitz determinants; however our Conjecture \ref{conjecture} also supposes that we can identify the individual eigenvalues that go to zero, going beyond the determinant]. The proof of this proposition follows closely parts of the proof of Widom's theorem that we turn to now.

\subsubsection{Widom's theorem}
In Ref.~\cite{Widom90}, the asymptotics of Toeplitz determinants of $z^\omega f_0(z)$ where $f_0(z)$ is continuous and piecewise $C^\infty$ but not $C^\infty$ are analysed. This means that there are finitely many points (singularities) $z_h=\rme^{\rmi \theta_h}$, and at each such point there is a finite $\alpha_h \in \mathbb{N}$  where the $\alpha_h$th derivative is discontinuous (and for all integers $k<\alpha_h$ the derivatives are continuous). This is a different definition of singularity compared to the one used in the main text (based on certain asymptotic expansions), but is related, and indeed the same picture emerges.
\begin{theorem}[Widom]\label{thm:Widom}
Suppose we have $f(z)=z^\omega f_0(z)$, where $f_0(z)$ is continous, non-zero and piecewise $C^{\infty}$ but not $C^{\infty}$ on the unit circle, and has winding number zero. Suppose that $f_0(z)$ has $m$ singularities at $z_1=\rme^{\rmi \theta_1}, \dots ,z_m=\rme^{\rmi \theta_m}$ with corresponding $\alpha_1,\dots,\alpha_m$. Then:
\begin{align}
D_N[z^\omega f_0(z)] = (-1)^{N \omega} D_N[f_0(z)] \left( N^{-e} \sum_{k_1,\dots k_m}c_{k_1,\dots k_m}\prod_{h=1}^m\left(G_h^{k_h}z_h^{N k_h} \right) +O(N^{-e-1}\Log(N)) \right).
\end{align}
The decay $e$ is given by $e=\min\{ \sum_{h=1}^m k_h^2+\alpha_h k_h: k_h \in \mathbb{Z}_+, \sum k_h = \lvert \omega \rvert\}$. The sum is taken over all $k_1 \dots k_m$ where this minimum is achieved,
the $c_{k_1,\dots k_m}$ are non-zero constants and  \begin{align}G_h = \rme^{\rmi \mathrm{sign}(\omega)\Log(f_0(z_h)}\lim_{\eps\rightarrow 0}\frac{f_0^{(\alpha_h)}(\rme^{\rmi(\theta_h+\eps)})-f_0^{(\alpha_h)}(\rme^{\rmi(\theta_h-\eps)})}{f_0(z_h)}.\end{align}
\end{theorem}
Defining $\Omega_{\theta_h} =\alpha_h+1$, the formula for $e$ is consistent with the singularity-filling picture as discussed in the main text; and in the case the minimum is unique will give the leading term (with non-zero coefficient for all $N$) as a rigorous result for the Toeplitz determinant. 
\subsection{Determinants of asymptotic expansions---Proof of Proposition \ref{prop:asymptotics}}\label{app:asymptotics}
In this section we find the asymptotics of Toeplitz determinants $D_n[z^{-N} l(z)]$ based on the asymptotic expansion of $l_N$. The key ideas are all based on Widom's proof of Theorem \ref{thm:Widom}. Let us first recall a lemma \footnote{There is a misprint in the statement of this lemma in \cite{Widom90}.}:
\begin{lemma}[Widom 1990]
Suppose we have a finite set of measures $\rmd \mu_n(s,t)$ and functions $\varphi_n(s), \psi_n(t)$ such that $\varphi_n(s)^j\psi_n(t)^k\in L_1(\rmd \mu_n(s,t))$. Define the matrix $M$ by:
\begin{align}
M_{jk} = \sum_{n} \int \varphi_n(s)^j\psi_n(t)^k\rmd \mu_n(s,t) \qquad  j,k=0,\dots,r-1,\label{Mjkdef}
\end{align}
then:
\begin{align}
\det(M) = \frac{1}{r!}\sum_{n_0,\dots n_{r-1}} \int \prod_{j>k}\Big(\left(\varphi_{n_j}(s_j)-\varphi_{n_k}(s_k)\right)\left(\psi_{n_j}(t_j)-\psi_{n_k}(t_k)\right)\Big) \prod_{l=0}^{r-1}\rmd \mu_{n_l}(s_l,t_l),
\end{align}
where the sum is over all $r$-tuples $(n_0,\dots n_{r-1})$.
\end{lemma}
\subsubsection{One singularity}
Suppose that we have a function $l(z)$ with a single singularity at $z=1$. By that, we mean that there is an asymptotic expansion of the form:
\begin{align}
l_N = \sum_{p \geq \alpha+1} a_p N^{-p}.  \label{lNexpansion}
\end{align}
Recall that for winding number $\omega<0$ we are interested in the determinant of the matrix $\tilde{M}$, where:
\begin{align}
\tilde M_{jk} = l_{N+j-k} \qquad j,k = 1, \dots, \lvert \omega\rvert.
\end{align}
Following Widom, to determine the asymptotics of this determinant to order $O(N^{-N_0})$ we can keep only finitely many terms of \eqref{lNexpansion} (e.g. take up to $p=N_0$). Then we can write the matrix elements of $M_{jk}$ as the finite sum:
\begin{align}
l_{N+j-k} = \sum_{p \geq \alpha+1} a_p (N+j-k)^{-p}=\sum_{p\geq\alpha+1}   \frac{a_p}{\Gamma(p)} \int_0^\infty \rme^{-Nt} \rme^{(k-j)t} t^{p-1}\rmd t.
\end{align}
This is of the form \eqref{Mjkdef}, with a sum over $p$
and 
\begin{align}
    \varphi_p(t) &= \rme^{-t} \qquad \psi_p(t)=\rme^t\nonumber\\
    \rmd \mu_{p}(s,t)&= \delta(s-t) \frac{a_p}{\Gamma(p)}\rme^{-Nt}t^{p-1}\rmd s\rmd t. 
\end{align}
Hence, defining $r=\lvert\omega\rvert$, we can write:
\begin{align}
\det(\tilde{M}) = \frac{1}{r!}\sum_{p_0,\dots p_{r-1}} \int \prod_{j>k}\Big(\left(\rme^{-t_j}-\rme^{-t_k}\right)\left(\rme^{t_j}-\rme^{t_k}\right)\Big)\rme^{-N \sum_{l=0}^{r-1}t_l} \prod_{l=0}^{r-1}\frac{a_{p_l}}{\Gamma(p_l)}t_l^{p_l-1}\rmd t_l .
\end{align}
Let us then rescale $t_j\rightarrow t_j/N$.
\begin{align}
\det(\tilde{M}) &=\frac{ N^{-\sum_l p_l} }{r!}\sum_{p_0,\dots, p_{r-1}} \int \prod_{j>k}\Big(\left(\rme^{-t_j/N}-\rme^{-t_k/N}\right)\left(\rme^{t_j/N}-\rme^{t_k/N}\right)\Big)\rme^{-\sum_{l=0}^{r-1}t_l} \prod_{l=0}^{r-1}\frac{a_{p_l}}{\Gamma(p_l)}t_l^{p_l-1}\rmd t_l\\
&=\frac{ N^{-\sum_l p_l} }{r!}\sum_{p_0,\dots, p_{r-1}} \int \prod_{j>k}\left(-N^{-2} (t_k-t_j + O(1/N))^2\right)\rme^{-\sum_{l=0}^{r-1}t_l} \prod_{l=0}^{r-1}\frac{a_{p_l}}{\Gamma(p_l)}t_l^{p_l-1}\rmd t_l\\
&=\frac{(-1)^{\frac{r^2-r}{2}} N^{-r^2+r-\sum_l p_l} }{r!}\sum_{p_0,\dots, p_{r-1}} \int \prod_{j>k}\left((t_k-t_j + O(1/N))^2\right)\rme^{-\sum_{l=0}^{r-1}t_l} \prod_{l=0}^{r-1}\frac{a_{p_l}}{\Gamma(p_l)}t_l^{p_l-1}\rmd t_l.
\end{align}
The leading order term is given by:
\begin{align}
\det(\tilde{M}) &=\frac{(-1)^{\frac{r^2-r}{2}} N^{-r^2+r-r(\alpha+1)} }{r!} \int \prod_{j>k}(t_k-t_j)^2\rme^{-\sum_{l=0}^{r-1}t_l} \prod_{l=0}^{r-1}\frac{a_{\alpha+1}}{\Gamma(\alpha+1)}t_l^{\alpha}\rmd t_l (1+o(1))\label{eq:asymptoticintegral}\\
&= a_{\alpha+1}^r{(-1)^{\frac{r^2-r}{2}} N^{-r^2-r\alpha} } \det(M_0) (1+o(1))
\end{align}
where $M_0$ is independent of $N$ and given by:
\begin{align}
(M_0)_{j,k} = \prod_{l=0}^{j+k-1}(\alpha+1+l) \qquad j,k=0,\dots,r-1.
\end{align}
In the last step, following \cite{Widom90}, we take the integral in \eqref{eq:asymptoticintegral} and use Lemma 1 to write it as the determinant of $M_0$.

\subsubsection{Multiple singularities}
Now suppose that we have:
\begin{align}
l_N &= \sum_{n=1}^m \sum_{p \geq \alpha_n+1} \rme^{\rmi N\theta_n } a_{n,p} N^{-p} \label{lNexpansion2}\\
l_{N+j-k} &= \sum_{(n,p)} \rme^{\rmi (N+j-k)\theta_n }\frac{a_{(n,p)}}{\Gamma(p)} \int_0^\infty \rme^{-Nt} \rme^{(k-j)t} t^{p-1}\rmd t
\end{align}
where $n$ labels the $m$ singularities on the unit circle. As before, we can keep only finitely many of the terms in the sum over $p$, so we have a finite sum over pairs $(n,p)$ with $p\geq\alpha_n+1$. This is of the form \eqref{Mjkdef} with
\begin{align}
    \varphi_{(n,p)}(t) &= \rme^{\rmi \theta_n}\rme^{-t} \qquad \psi_{(n,p)}(t)=\rme^{-\rmi \theta_n}\rme^t\nonumber\\
    \rmd \mu_{(n,p)}(s,t)&= \delta(s-t) \frac{a_{(n,p)}}{\Gamma(p)}\rme^{\rmi N \theta_n}\rme^{-Nt}t^{p-1}\rmd s\rmd t. 
\end{align}
Hence, defining $r=\lvert\omega\rvert$, we can write:
\begin{align}
\det(\tilde{M})& = \frac{1}{r!}\sum_{(n_0,p_0),\dots ,(n_{r-1},p_{r-1})} \int \prod_{j>k}\Big(\left(\rme^{\rmi \theta_{n_j}}\rme^{-t_j}-\rme^{\rmi \theta_{n_k}}\rme^{-t_k}\right)\left(\rme^{-\rmi \theta_{n_j}}\rme^{t_j}-\rme^{-\rmi \theta_{n_k}}\rme^{t_k}\right)\Big)\nonumber\\
&\qquad\qquad\times\rme^{-N \sum_{l=0}^{r-1}t_l} \prod_{l=0}^{r-1}\frac{\rme^{\rmi N \theta_{n_l}}a_{(n_l,p_l)}}{\Gamma(p_l)}t_l^{p_l-1}\rmd t_l.
\end{align}
As before we rescale $t_j\rightarrow t_j/N$. Note that for $n_j=n_k$ we have:\begin{align}
\left(\rme^{\rmi \theta_{n_j}}\rme^{-t_j/N}-\rme^{\rmi \theta_{n_k}}\rme^{-t_k/N}\right)\left(\rme^{-\rmi \theta_{n_j}}\rme^{t_j/N}-\rme^{-\rmi \theta_{n_k}}\rme^{t_k/N}\right) = -N^{-2}\left((t_k-t_j)+O(1/N)\right)^2 
\end{align}
while for $n_j\neq n_k$:
\begin{align}
\left(\rme^{\rmi \theta_{n_j}}\rme^{-t_j/N}-\rme^{\rmi \theta_{n_k}}\rme^{-t_k/N}\right)\left(\rme^{-\rmi \theta_{n_j}}\rme^{t_j/N}-\rme^{-\rmi \theta_{n_k}}\rme^{t_k/N}\right) &= (\rme^{\rmi \theta_{n_j}}-\rme^{\rmi \theta_{n_k}})(\rme^{-\rmi \theta_{n_j}}-\rme^{-\rmi \theta_{n_k}})+O(1/N)\nonumber\\
&= \underbrace{\left(2\sin\left(\frac{\theta_{n_j}-\theta_{n_k}}{2}\right)\right)^2}_{s(n_j,n_k)>0} +O(1/N).
\end{align}
Then, for each choice of $\{n_j\}$ we have a dominant term:
\begin{align}
&   \frac{N^{-\sum_{l=0}^{r-1}{(\alpha_{n_l}+1)}}\rme^{\rmi N \sum_{l=0}^{r-1}\theta_{n_l}}}{r!} \int \prod_{j>k, n_j=n_k} \left(-N^{-2}(t_k-t_j)^2\right)  \prod_{j>k, n_j\neq n_k} s(n_j,n_k)
\rme^{-\sum_{l=0}^{r-1}t_l} \prod_{l=0}^{r-1}\frac{a_{(n_l,\alpha_{n_l}+1)}}{\Gamma(\alpha_{n_l}+1)}t_l^{\alpha_{n_l}}\rmd t_l&\\
&= (-1)^{x}\frac{N^{-\sum_{l=0}^{r-1}{(\alpha_{n_l}+1)}-2x}\rme^{\rmi N \sum_{l=0}^{r-1}\theta_{n_l}}}{r!}\left(\prod_{l=0}^{r-1}{a_{(n_l,\alpha_{n_l}+1)}}\right) \nonumber\\&\qquad\times\underbrace{\int \prod_{j>k, n_j=n_k} \left(t_k-t_j\right)^2 \times \prod_{j>k, n_j\neq n_k} s(n_j,n_k)
\rme^{-\sum_{l=0}^{r-1}t_l} \prod_{l=0}^{r-1}\frac{t_l^{\alpha_{n_l}}}{\Gamma(\alpha_{n_l}+1)}\rmd t_l}_{c(\{n_l\})>0}.
\end{align}
where we define $x$ to be the number of pairs $(j,k)$ such that $j>k$ and $n_j=n_k$, and the integral $c(\{n_l\})$ is positive. Now, let us find the dominant term(s) among these choices of $\{n_j\}$. Recall that $n_0,\dots n_{r-1}$ will correspond to $r$ choices of the $m$ singularities. Let us suppose that the the singularity corresponding to $\alpha_n$ has filling $k_n$, then overall we have $\sum_n k_n = r$. We also have that the number of $(j,k)$ such that $n_j=n_k =n$ is given by $k_n(k_n-1)/2$. Hence, $2x=\sum_n k_n(k_n-1)$, and moreover:
\begin{align}
N^{-\sum_{l=0}^{r-1}{(\alpha_{n_l}+1)}-2x}=N^{-\left(\sum_{n=1}^m{k_n^2 + k_n\alpha_{n}}\right)}.
\end{align}
Hence, we have that the dominant terms have asymptotic behaviour $N^{-e}$ where:
\begin{align}
e=\min\{ \sum_{n=1}^m k_n^2+\alpha_n k_n \quad \mathrm{for~}  k_n \in \mathbb{Z}_+, \sum_{n=1}^m k_n = r=\lvert \omega \rvert\}.
\end{align}
Note that we can write $c(\{n_l\})=c_{k_1,\dots k_m}$ since we can order the $t_k$ as we like. As in Widom's result, it is possible that the different dominant contributions could cancel, so that the dominant asymptotic term is not $N^{-e}$. We have the correct asymptotics if the minimum is unique, for example. This analysis for $l(z)$ could be repeated identically for $m(z)$, now the behaviour will depend on $\alpha_n$ that characterise the singularities of $m(z)$. If either $l(z)$ or $m(z)$ is analytic on an annulus containing the unit circle, then we should include in the expansion those terms coming from the nearest singularity to the unit circle.

Finally, putting $\Omega_{\theta_n} = \alpha_n+1$ we have Proposition \ref{prop:asymptotics}, and this agrees with the singularity-filling picture of Conjecture \ref{conjecture}.
\section{Proof of Theorem \ref{thm3}} \label{app:order}
Recall that $\langle \mathcal{O}_\kappa (1)\mathcal{O}_\kappa (N)\rangle$ is a Toeplitz determinant generated by $F_\kappa(z)= z^{-\kappa} f(z)/\lvert f(z)\rvert$ (see \cite{Jones19} for details). This means that $F_\omega(z)$ has winding number zero, and has a continuous logarithm $W(z)$. We assume $f(z)$ is $\alpha$-decaying, and write $\alpha = 1+\beta$. Using the result of Ref.~\cite{Gong22}, we have that $f(z)/|f(z)|$ is $(\alpha-\eps)$-decaying for any $\eps>0$, so put $\eps=\beta/2$. Then  $F_\omega(z)$ is $\gamma$-decaying for some $\gamma>1$.
This in turn means that if we define $F_n= (F_\omega(z))_n$, then $\sum_{n=-\infty}^\infty |n| |F_n|^2 < \infty$. Hence Szeg\"{o}'s theorem (as stated in \cite{Hartwig69}) applies to $F_\omega(z)=\rme^{W(z)}$, giving us that:
\begin{align}
\langle \mathcal{O}_\omega (1)\mathcal{O}_\omega (N)\rangle=D_{N-1}[F_\omega(z)] = \rme^{\rmi \pi s (N-1)}\rme^{\sum_{k=0}^\infty k W_k W_{-k}}(1+o(1)).
\end{align}
The sign of $F_\omega(1)$ is equal to $(-1)^s$.
Theorem \ref{thm1}
The condition that $F_\omega(z)$ is $\gamma$-decaying for $\gamma>1$ also implies that if we write $\tilde{F}_n = \max_{m\geq n}\{|F_m|, |F_{-m}|\}$, then $\sum_{n=-\infty}^\infty \tilde{F}_n <\infty$. Then we can use Theorem 4 of Hartwig and Fisher \cite{Hartwig69}: this gives us that for all $0\neq \delta \in \mathbb{Z}$, the correlator $\langle \mathcal{O}_{\omega+\delta} (1)\mathcal{O}_{\omega+\delta} (N)\rangle= D_{N-1}[z^{-\delta} F_\omega(z)]=o(1)$ as $N\rightarrow \infty$. This completes the proof of Theorem \ref{thm3}.

\section{Gapless models}\label{app:gapless}
\subsection{Comments}
We first note that the argument given in the main text showing that $f(z)=(z-1)f_g(z)$ (a direct transition between models with non-trivial winding) has edge mode(s), is consistent with the general picture that we expect that non-trivial boundary physics can occur at transitions between non-trivial models (i.e., where the critical point cannot be perturbed into the trivial phase) \cite{Verresen18,Verresen20,Verresen21}. 

We claim that gapless short-range models are generically of a form similar to the one analysed in the main text: i.e., a polynomial with zeros on the unit circle, multiplied by a function that corresponds to a gapped Hamiltonian. 
Then we can use Theorem \ref{thm1} to prove the existence of edge modes. By short-range we include both the finite-range case and the case where $f_c(z)$ is meromorphic with no poles on the unit circle (and hence has exponentially decaying Fourier coefficients). To see this is a general form we use the Fundamental Theorem of Algebra, and the Weierstrass Factorisation Theorem \cite{Lang03}. Note also that degenerate zeros are straightforwardly accounted for, but we no longer have linearly dispersing critical modes.
\subsection{Example}
We can consider direct transitions between two of the chains \eqref{eq:example2}; corresponding to models with $f(z) = z^\kappa(1+ a \mathrm{Li}_\alpha(z))$ for different constants $a$ and values of $\alpha>2$. For $\kappa=0$ (and appropriate $a$), these models have $\omega=0$ and we can write an interpolation between $\omega=1$ and $\omega=0$ by:
\begin{align}
f(z) =  \lambda z(1+a\mathrm{Li}_\beta(z))-(1-\lambda) \left(1+b\mathrm{Li}_\gamma(z)\right),
\end{align}
where we fix $b = a \frac{\zeta(\beta)}{\zeta(\gamma)}$. We see that there is a gapless mode at $k=0$ for $\lambda =1/2$; indeed, $f(z=1) \neq 0$, except at this critical point. We now show that, for $\lambda=1/2$, $f(z) = (z-1)f_g(z)$ for a gapped model $f_g(z)$ that is $\alpha$-decaying with $\alpha>1$.

Firstly, $f(z)$ and $(1-z)^{-1}$ are analytic inside the unit circle, hence, for $\lvert z \rvert <1$, we can write:
\begin{align}
f_g(z) = - f(z)(1-z)^{-1} = 1 - a \sum_{n=1}^\infty \left(H_n^{(\beta)}z^{n+1} -\frac{\zeta(\beta)}{\zeta(\gamma)}H_n^{(\gamma)}z^n \right)= 1-a\sum_{n=1}^\infty \underbrace{\left(H_{n-1}^{(\beta)} -\frac{\zeta(\beta)}{\zeta(\gamma)}H_n^{(\gamma)} \right)}_{a_n}z^n,
\end{align}
where $H_n^{(\alpha)}= \sum_{m=1}^n m^{-\alpha}$ is the $n$th harmonic number of order $r$. Using the Euler-Maclaurin formula \cite[Eq.~2.10.7]{NIST:DLMF} we have the following expansion for this harmonic number: 
\begin{align}
H_n^{(\alpha)} = \zeta(\alpha) - \frac{1}{\alpha-1}n^{-(\alpha-1)}\sum_{s=0}^\infty \left(\begin{array}{c}1-\alpha \\s\end{array}\right) \frac{B_s}{n^s},  \label{eq:harmonic}
\end{align}
where $B_s$ is the $s$th Bernoulli number.
Thus
\begin{align}    
a_n =  \frac{\zeta(\beta)}{\zeta(\gamma)}\frac{1}{\gamma-1}n^{-(\gamma-1)}\sum_{s=0}^\infty \left(\begin{array}{c}1-\gamma \\s\end{array}\right) \frac{B_s}{n^s}- \frac{1}{\beta-1}n^{-(\beta-1)}\sum_{s=0}^\infty \left(\begin{array}{c}1-\beta \\s\end{array}\right) \frac{B_s}{n^s}.
\end{align}
For large enough $s$, truncating the sums give us a remainder that is approximated by the first neglected term \cite{NIST:DLMF}, and so we conclude that $f_g(z)$ is $\delta$-decaying, for $\delta=\min(\beta-1,\gamma-1)$. This implies the absolute convergence of $f_g(z)$ on the unit circle. Hence, we have some finite range of $a$ where $f_g(z)=1+h(z)$ such that $\lvert h(z)\rvert <1 $, and so $f_g(z)$ has winding number zero, and moreover is non-vanishing on the unit circle.

We can then use the reasoning as in the main text to argue that we have edge modes at critical points of the form $f_c(z) = z^\kappa f(z) $ for different values of $\kappa$. For $\kappa>0$, by considering the Hamiltonian directly, we can see that $f(z) = (z-1)z^\kappa f_g(z)$ has $\kappa$ exactly-localised edge modes. For $\kappa<-2$ we see non-trivial critical edge modes. Indeed, we have the Wiener-Hopf decomposition $b_+(z) = f_g(z)$, and $b_-(z)=1$. The model $f(z) = (1-1/z) z^{-(\lvert \kappa\rvert-1)} f_g(z)$ will have $\lvert \kappa\rvert-1$ edge modes that are shared by the gapped models $z^{-(\lvert \kappa\rvert-1)} f_g(z)$ and $z^{-(\lvert \kappa\rvert-2)} f_g(z)$; so by linearity we have these edge modes also for the critical Hamiltonian. Using Theorem \ref{thm1}, we can compute the non-trivial localisation of these edge modes by taking the Fourier coefficients of $1/f_g(z)$. 

\section{The AIII class}\label{app:AIII}
Throughout this work we have focused on the translation-invariant BDI class of free-fermion chains with integer topological invariant, with Hamiltonian given in the main text. Another such tenfold way class is AIII \cite{Altland97,Ryu10}, which has a realisation in a model with number-conserving complex fermions on two sublattices $A$ and $B$ \cite{Balabanov21,Jones21a}. The model has a sublattice symmetry forbidding hopping on the same lattice and is given by:
\begin{align}
H_{\textrm{AIII}}=\sum_{n,\alpha}\tau^{\vphantom\dagger}_{\alpha}c^\dagger_{B,n}c^{\vphantom\dagger}_{A,n+\alpha} + \overline{\tau}^{\vphantom\dagger}_{\alpha}c^\dagger_{A,n}c^{\vphantom\dagger}_{B,n-\alpha}.
\end{align}
Similar to BDI we can solve by Fourier transform followed by a rotation, summarised by $f(z) = \sum_{n} \tau_n z^n$, where now $\tau_n \in \mathbb{C}$. The model is gapped when $f(z)$ does not vanish on the unit circle, and in that case the winding number is well defined for absolutely-summable $\tau_n$. Moreover, the absolute-summability again implies the existence of a well-behaved Wiener-Hopf decomposition $f(z) = z^\omega b_+(z) b_-(z)$ (as before, we fix $V_0=0$). For a complex function $h(z)= \sum_\alpha h_\alpha z^\alpha$ we also define the notation $\overline{h}(z)= \sum_\alpha \overline{h}_\alpha z^\alpha$.
The results for the BDI class carry over straightforwardly, the main difference is in the physical interpretation. We will sketch the key points in this section.
\subsection{Edge modes}
An edge mode in AIII on the $A$-sublattice is a complex fermion of the form $c_{A,L}^\dagger = \sum_{n\geq 0} g_n^{\vphantom \dagger} c_{A,n}^\dagger$, that satisfies $[H_{\textrm{AIII}},c_{A,L}^\dagger ]=0$.
Evaluating this commutator as in the proof of Theorem \ref{thm1} we reach
\begin{align}
\sum_{n,m\geq0} \tau_{m-n}^{\vphantom\dagger} g_m^{\vphantom\dagger} c^\dagger_{B,n} =0,
\end{align}
which is equivalent to $\sum_{m\geq 0} \tilde\tau_{n-m} g_m = 0$ for all $n$, where we define $\tilde\tau_{n-m} =\tau_{m-n}$. Using the Wiener-Hopf sum equation results, we thus have $\nu>0$ $A$-sublattice edge modes if $c(z) = \sum_{\alpha} \tilde\tau_\alpha z^\alpha = {f}(1/z)$ has winding number $-\nu$. This is equivalent to $f(z)$ having winding number $\omega =\nu>0$. The coefficients $g_n$ of the wavefunction are Fourier coefficients of $b_-(1/z)^{-1}$.

If $f(z)$ has $\omega<0$ then, through the same steps, we will have $\lvert \omega \rvert$ $B$-sublattice edge modes of the form  $c_{B,L}^\dagger = \sum_{n\geq 0} g_n^{\vphantom \dagger} c_{B,n}^\dagger$. The relevant function in the Wiener-Hopf sum equation is $c(z) =  \overline{f}(z)$, which has the same winding as $f(z)$. However, this complex conjugation of the coefficients means the $g_n$ are given by Fourier coefficients of $\overline{b}_+(z)^{-1}$.
\subsection{Splittings}
In this class the single-particle Hamiltonian is the block Toeplitz matrix generated by
\begin{align}
\Phi(z) = \left(\begin{array}{cc}0 & f(z) \\-\overline{f}(1/z) & 0\end{array}\right).
\end{align}
Then $\prod_{j=1}^L(-\eps_j^2)=D_L[\Phi(z)] = (-1)^L D_L[f(z)]D_L[\overline{f}(1/z)]=(-1)^L \lvert D_L[f(z)]\rvert^2 $.

The same Toeplitz determinant theory will apply here. Hence, given the various assumptions as discussed in the BDI case, we will have a singularity-filling picture for finite-size splitting as before. 

\subsection{String-order parameters}
As proved in Ref.~\cite{Jones21b}, there are string correlators $\tilde{\mathcal{O}}_{\kappa}$ in the finite-range AIII class that can be evaluated as the Toeplitz determinant \begin{align}
\langle \tilde{\mathcal{O}}_\kappa(1) \tilde{\mathcal{O}}_\kappa(N)\rangle= 
    \Big\lvert D_{N-1} \left[z^{-\kappa} {f(z)}/{|f(z)|}\right] \Big\rvert^2.
\end{align}
$\tilde{\mathcal{O}}_{\kappa}$ are decorated parity strings:
\begin{align}
\tilde{\mathcal{O}}_0(n) &= \exp\left(\sum_{m=1}^{n-1}\rmi\pi\left(c^\dagger_{A,m}c^{\vphantom \dagger}_{A,m}+c^\dagger_{B,m}c^{\vphantom \dagger}_{B,m}\right)\right) \\
\tilde{\mathcal{O}}_\kappa(n) &=\exp\left(\sum_{m=1}^{n-1}\rmi\pi\left(c^\dagger_{A,m}c^{\vphantom \dagger}_{A,m}+c^\dagger_{B,m}c^{\vphantom \dagger}_{B,m}\right)\right)\prod_{j=n}^{n+\kappa-1}(1-2c^\dagger_{A,j}c^{\vphantom \dagger}_{A,j})\nonumber \qquad \textrm{for~}\kappa>0\\
\tilde{\mathcal{O}}_{\kappa}(n) &= \exp\left(\sum_{m=1}^{n-1}\rmi\pi\left(c^\dagger_{A,m}c^{\vphantom \dagger}_{A,m}+c^\dagger_{B,m}c^{\vphantom \dagger}_{B,m}\right)\right)\prod_{j=n}^{n+\lvert\kappa\rvert-1}(1-2c^\dagger_{B,j}c^{\vphantom \dagger}_{B,j})\nonumber\qquad \textrm{for~} \kappa<0 .
\end{align}
Thus, assuming $\alpha$-decaying $\tau_n$, the proof of Theorem \ref{thm3} carries over (again, this relies on the clustering result of Ref.~\cite{Gong22}). Hence, the string-orders continue to act as order-parameters even in the long-range ($\alpha>1$-decaying) AIII class.
\end{document}